\documentclass[aps, prd, amsmath, floats,floatfix, twocolumn,superscriptaddress, nofootinbib]{revtex4}
\usepackage{amsmath} 
\usepackage{verbatim} 
\usepackage{mathrsfs}
\usepackage{amsfonts}
\usepackage{dsfont}
\usepackage{color}
\usepackage{graphicx}
\usepackage{subcaption}
\usepackage{booktabs}
\usepackage{multirow}

\newcommand{\be}{\begin{equation}}
\newcommand{\ee}{\end{equation}}
\newcommand{\bea}{\begin{eqnarray}}
\newcommand{\eea}{\end{eqnarray}}

\begin{document}

\title{Modeling dark matter halos with self-interacting fermions: A polytropic approach}
\author{Fabian Hernandez-Gutierrez}
\email{f.hernandezgutierrez@ugto.mx}
\affiliation{Division de Ciencias e Ingenier\'ias,  Universidad de Guanajuato, Campus Leon,
 C.P. 37150, Le\'on, Guanajuato, M\'exico.}
\author{J. Barranco}
\email{jbarranc@fisica.ugto.mx}
\affiliation{Division de Ciencias e Ingenier\'ias,  Universidad de Guanajuato, Campus Leon,
 C.P. 37150, Le\'on, Guanajuato, M\'exico.}

\date{\today}

\begin{abstract}

In this work we study the possibility of modeling the dark matter content in galaxies as a core-halo model consisting of self-gravitating, self-interacting fermions  by means of an effective polytropic equation of state. 
For the core of the halo, the dark matter fermions are degenerate, while for the halo we have considered two possibilities: the fermions have thermalized as a perfect fluid, or they will follow a standard cold dark matter Navarro-Frenk-White profile. The core density profile is obtained by solving the Tolman-Oppenheimer-Volkoff equations, and their properties are determined by the fermion mass, the central density and the interaction strength. The mass of the fermion and the strength of the fermion self-interaction is fixed by doing a $\chi^2$ analysis to fit that fit the rotational curves of Low Surface Brightness galaxies. 

It was found that the fermion mass should be in the range $155.10~\rm{eV}< m_{f} < 313.26~\rm{eV}$ and the interparticle strength in the range $1465.31 < y <6002.04$  at $68$ C.L. in order to reproduce the rotational curves adequately, in the case when the halo is modeled as a thermalized ideal gas. Similar values are obtained if the halo is modeled following a Navarro-Frenk-White case, namely $42.31 ~\rm{eV} < m_{f} <49.23 ~\rm{eV}$ and $71.43< y <  132.65$.

Once fixed the values of the mass of the fermion $m_f$ and the interaction strength $y$, we tested the core-halo model with data from the Milky Way, local dwarf spheroidal galaxies and the SPARC database. We have found good agreement between the data and the core-halo models, varying only one free parameter: the central density. Thus a single fermion can fit hundreds of galaxies. Nevertheless, the dark matter halo surface density relation or the halo total mass and radius depend strongly on the model for the halo. 
\end{abstract}
\maketitle

\section{Introduction}

For most cosmological data the paradigm of Lambda Cold Dark Matter ($\Lambda$CDM) is suitable and reproduces the observations well enough \cite{Planck2018}. The area in which this model has inconsistencies is the galactic scales where the predicted and universal Navarro-Frenk-White (NFW) dark matter profile \cite{1997Navarro} and observations do not agree. Some of the well-documented problems with $\Lambda$CDM at small scales are: the Cusp/Core problem \cite{Bullock:2017xww}, the Missing Satellite problem \cite{Klypin:1999uc}, and the Too-Big-to-Fail problem \cite{Boylan-kolchin2011}. Lesser known problems include: the satellite plane problem \cite{Pawlowski:2018sys}, the discrepancy of the Baryonic Tully-Fisher relation (BTFR), inferred from simulations compared to observations \cite{McGaugh:2012}, and the observed survival of a globular cluster for a Hubble time which is not consistent if galactic halos are cusped, as CDM predicts \cite{Kleyna:2003zt}. Some works that tackle those issues at galactic scale have proposed adding baryonic interaction to the CDM model to deal with them \cite{Simon2007, El_Zant_2001, Navarro_1996, Zolotov_2012,Brooks_2014} but there is no general consensus that baryons will solve all those problems.\\

Cold dark matter (CDM) is effectively modeled as a perfect fluid with an equation of state (EoS) $p \simeq 0$. In the simplicity of this description underlies the difficulty of extracting the properties of the particle associated with dark matter. Furthermore, it is possible that the small scale problems of CDM relies on this pressure-less hypothesis. Motivated by the today's age of precision cosmology \cite{Turner:2022gvw} and the necessity to find a solution to the galactic scale problems of $\Lambda$CDM, it is time to move beyond the pressure-less CDM paradigm.
Early attempts to test an equation of state for dark matter at cosmological scales have been done in
\cite{Muller:2004yb,Calabrese:2009zza,Kumar:2012gr,Armendariz-Picon:2013jej,Xu:2013mqe}. 
Those works have constrained a barotropic EOS for DM $p=\omega \rho$ with $-4\times10^{-5}<\omega< 1.453\times10^{-3}$ \cite{Xu:2013mqe}. 
Stronger constraints are obtained at galactic level \cite{Serra:2011jh,Barranco:2013wy,Acena:2021wjx,Boshkayev:2021wns} by fitting the rotational velocities in spiral galaxies. It was found that an effective pressure within the limits obtained by cosmological studies can be sufficient to explain the rotational curve data and the cores of galaxies. \\

Another possibility considers that the intrinsic nature of the dark matter particle could solve some of the problems. By the “nature of dark matter,” we refer to the essential and irreplaceable properties of DM particles: its mass, spin, and interactions with other particles or itself; information still unknown for dark matter. These attributes influence the way galactic-scale structures are formed.  It is of special interest the spin of the particle as it alters the their statistical properties and can lead to different halo shapes and mass distributions.  The two main options studied in relation with the spin of dark matter are the fermionic spin $1/2$ particles, like the WIMP and sterile neutrinos, that obey Fermi-Dirac statistics and spin-$0$ candidates that follow Bose-Einstein statistics, like axions and fuzzy dark matter. \\

This two possibilities converge when it is realized that both, ultra-light fermions and bosons have an intrinsic pressure. Theories like Fuzzy cold dark try to solve problems in the Cold Dark Matter (CDM) paradigm, 
considering the quantum pressure generated by the Heisenberg uncertainty principle  \cite{Hu2000}.  
For candidates with spin $1/2$ the pressure sustaining the galaxy from collapse would come from both the Heisenberg uncertainty principle and the Pauli exclusion principle \cite{DKoester_1990,Domcke2015} , stabilizing it gravitationally due to fermionic quantum degeneration pressure and producing core profiles.\cite{Randall2017, deVega:2013ysa}. It has been shown that a Fermi degenerate region must exist in ``a physically relevant region'' \cite{Randall2017}, making the self gravitating system a natural quantum macroscopic object, which would make it necessary to consider quantum mechanical calculations to compute galaxy structures at kpc scales and below \cite{Destri2013}, and turning the classical $N$-body simulations not appropriate below a certain $r_{min}$, where quantum mechanical effects are essential, which could explain the issues of $N$-body simulations at galactic scales.\\

Fermionic dark matter and self gravitating objects made of fermions have been studied to different degrees of complexity, including transition phases and isothermal envelope \cite{chavanis:hal-00139138, chavanis:hal-01094150} . It is of special notice the model of Ruffini-Arguelles-Rueda in which halos are constructed by a maximum entropy principle and has been tested with different observations \cite{Arguelles:2023nlh}.\\

The simplest model of a fermionic dark matter model is a degenerate gas of fermions which only interacts with the standard model particles gravitationally.  This gas will be degenerate as long as conditions for the degeneracy to apply exist. Outside this region the system would then dilute to the classical isothermal equation of state \cite{Chavanis2022}, leading to what is commonly known as a ``core-halo" system.\\ 

In this work, by simultaneously analyzing the rotational curve of low surface galaxies, we found that it is not possible to model all the galaxies with a single species of fermion if it lacks self-interaction. It is found that each galaxy needs a different value of the mass of the fermion. 
In the context of dark matter with pressure, it is common to invoke self-interaction between the dark matter particles 
\cite{Carlson:1992fn,deLaix:1995vi,Firmani:2000ce,Spergel:1999mh,Vogelsberger:2012sa}. 
Self-interaction can mitigate the small scale problems of CDM by producing less dense cores due to elastic scattering.  
When self-interaction in the model is considered, collisional relaxation can lead to a thermodynamic equilibrium, resulting in a core profile with a polytropic shape \cite{Sanchez2021}. The self-interaction would lead to an isothermal envelope \cite{Chavanis2022}. \\ 

In general, it is difficult to combine microscopic phenomena with macroscopic objects like the dark matter halos. The mechanism generally used in the literature to include self-interaction and the model of dark matter halos consists in computing the radius where collision of dark matter particles can occur within a Hubble time. From that point inwards the halo is considered as a gas with barotropic equation of state.\\

In the case of dark matter with self-interaction \cite{Kamada:2016euw}, it was found that it is possible to reproduce multiple rotation curves considering a fixed $\sigma/m$ but does not explain the nature of the particle. Other possibilities consider modification to the Boltzmann equation in N-body simulations, although the size of the particles in N-body are macroscopic objects with masses of the order of $10^{5}M_\odot$.\\

In this work, we aim to test if self-interacting fermions can form dark matter halos with a core that can adequately reproduce the rotation curves of different galaxies with specific values of mass and self-interaction parameter. The self-interacting fermions are described with a well motivated equation of state which will be described in section \ref{eos}. As our focus is on the equation of state, our approach to include self-interaction differs from previous analysis, while still explaining a large number of observables. We will examine a core-halo model in which the core consists of a degenerate gas of fermions and the halo is a perfect fluid with an equation of state $p=\omega\rho$, to test this core-halo model we will also compare it with a case where the core is composed of the same degenerate fermion and the halo is the standard NFW halo. In this model, the fermions interact only gravitationally with the standard model and may also exhibit self-interaction.\\

The velocity profile is obtained by solving the Tolman-Oppenheimer-Volkoff system that models the degenerate core of the object, and computing the mass profile for each type of halo. This is presented in section \ref{model}. Taking this system, we fit the rotational velocities of some low-surface brightness (LSB) galaxies and extract the best fit for the mass of the fermion and the self-interaction strength constant. The details of the fit are described in section \ref{lsb}. Next, we show that with the values for $m_f$ and $y$ obtained with the LSB data, it is possible to fit the rotational velocity of the Milky Way. Furthermore, with the same values, we show that the 175 galaxies reported in the SPARC catalog can be fitted by only varying the central density of the dark matter halo. Thus, we conclude that the diversity of galaxies can be well explained by a single fermion with mass $m_f\simeq 45$ eV  or $m_f\simeq 300$ eV and a strong self-interacting coupling constant $y \simeq 96$ or $y \simeq 5\times10^3$, depending on the halo type.\\

In the present article we have only considered observations at galactic scales. We will leave the study of dark matter fermions with self-interaction at cosmological scales, via an effective polytropic equation of state, for a future work. 
Some conclusions and a discussion of the possible signature in the morphology of galaxies for this dark matter particle are drawn in section \ref{conclusions}.

\section{Equation of state for degenerate fermions with self-interaction }\label{eos}
{\bf }
In the degenerate limit there are explicit equations that describe the energy density $\rho$ and pressure $p$, which depend on $z$, the dimensionless Fermi momentum \cite{Chandrasekhar1939}. 
 For the  interaction between fermions, relativistic mean field (RMF)\cite{Serot1992} approach has been
been applied to construct the equation of
state (EOS) for neutron stars and supernovae \cite{Shen:1998gq} or neutralino stars \cite{Ren:2006tr}. 
In our case, following \cite{PhysRevD} and \cite{Stiele:2010xz}, we effectively introduce a contact term to represent the inter-particle interaction between dark matter fermions \cite{Stiele:2010xz}. the Langrangian for this interaction is 
\begin{equation}
\mathcal{L}_{int}=g\bar{\chi}\gamma_{\mu}\chi\mathbf{V}^{\mu} ,
\end{equation}
where the interaction strength is given by $y=gm_f /m_I$ \cite{PhysRevD}, where $m_I$ is the energy scale of the interaction, $m_f$ is the fermion mass and $g$ is the coupling constant.

\begin{widetext}
\begin{equation}
\rho=\frac{m_{f}^{4}}{8\pi^{2}}\left[(2z^{3}+z)(1+z^{2})^{1/2} - \sinh^{-1}(z)\right]+\left(\frac{m_{f}^{2}}{3\pi^{2}}\right)^{2}y^{2}z^{6},
\label{density}
\end{equation}

\begin{equation}
    p=\frac{m_{f}^{4}}{24\pi^{2}}\left[(2z^{3}-3z)(1+z^{2})^{1/2} + 3\sinh^{-1}(z)\right]+\left(\frac{m_{f}^{2}}{3\pi^{2}}\right)^{2}y^{2}z^{6}
    \label{pressure}
\end{equation}

\end{widetext}

As long as the temperature is below the temperature of degeneracy
\cite{Carroll:1009754}

\begin{equation}
    T<T_{deg}=\frac{\hbar^{2}}{2 m_{f}K_{B}}\left(\frac{3 \pi^2\rho(r)}{m_{f}}\right)^{2/3}.
\end{equation}

\begin{figure}
    \centering
    \includegraphics[width=\linewidth]{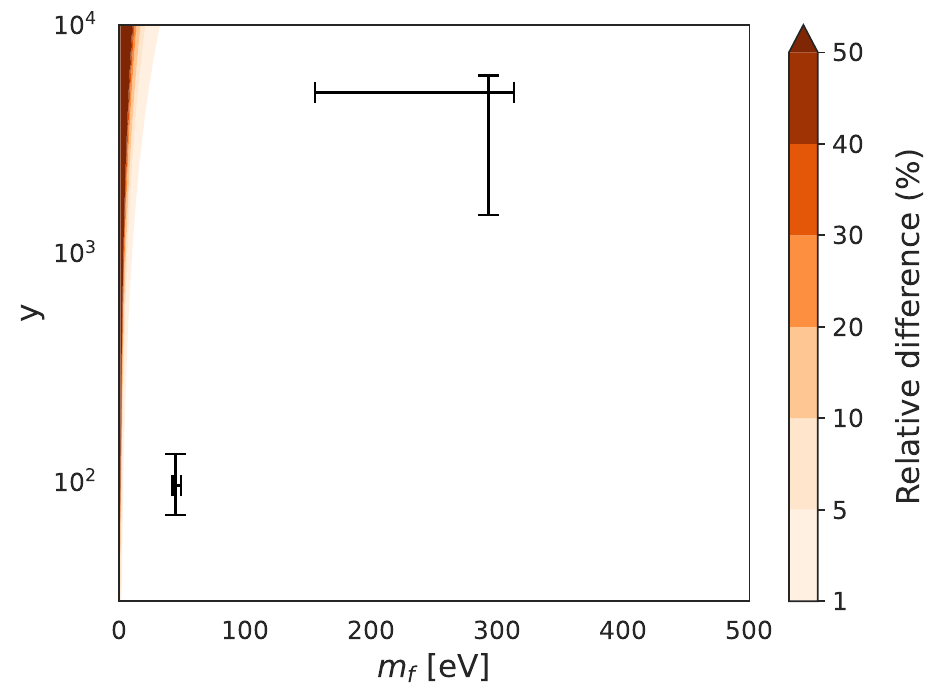}
    \caption{Relative difference (in \%) between the equation of state and its approximation as a function of fermion mass $m_f$ and interaction strength $y$. The color map shows the contour levels of the relative difference. The black error bars indicate the combined best-fit values (with 1$\sigma$ uncertainties) obtained from fits to six LSB galaxies, for the NFW halo (bottom) and the perfect-fluid halo (top). The green rectangles correspond to the 1$\sigma$ confidence regions of the individual galaxy fits.}
    \label{relative_difference}
\end{figure}

 we can take the non-relativistic limit were $z \ll 1$, that is when the mass of the fermion is much greater than the fermi momentum $m_{f} \gg k_{F}$ , and find that the equation of state for a free gas of fermions in the non relativistic case is 

\begin{equation}
    p=\frac{(3\pi^{2})^{2/3}}{5}\frac{\rho^{5/3}}{m_{f}^{8/3}} + \frac{y^{2}\rho^{2}}{m_{f}^{4}}
\label{polytrope}
\end{equation}

To corroborate the validity of this approximation we compared the polytropic equation of state with the full expressions for density and pressure given by Eqs.~(\ref{density})–(\ref{pressure}). As shown in Fig.~\ref{relative_difference}, the polytropic equation reproduces the exact equations to high accuracy, with the results presented in subsequent sections lying within a relative difference of less than 1\% across the explored parameter space. This supports the use of the polytropic form for practical purposes in the analysis.

\section{Self-gravitating ultralight fermions as Dark matter halos}\label{model}
\begin{figure*}
    \includegraphics[width=0.45\linewidth]{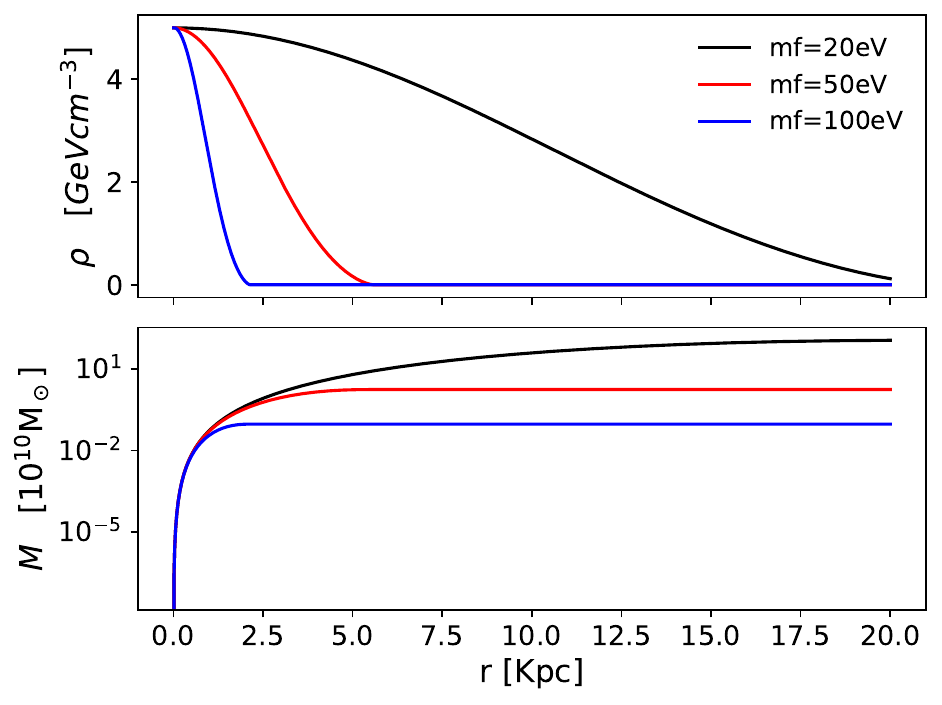}
    \includegraphics[width=0.45\linewidth]{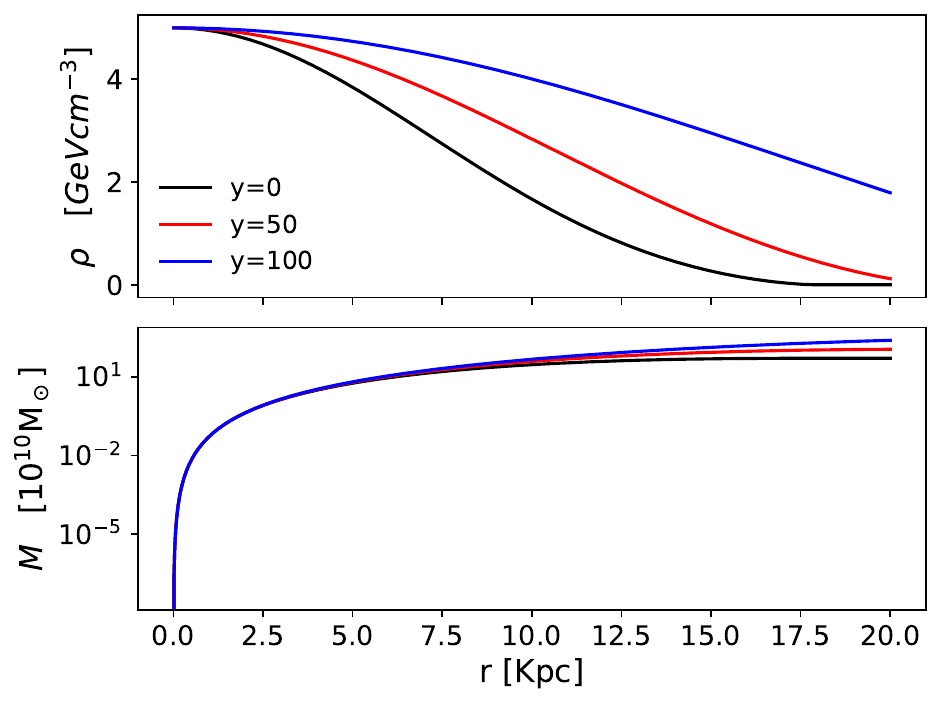}
    \caption{Density and mass profiles of the self-gravitating system composed of self-interacting degenerate fermions. The left panel shows the changes produced by different values of the fermion mass, with fixed values of $y=100$, $\rho_{0}=30\ GeV/cm^{3}$. The right panel shows the influence of the self-interaction parameter $y$, with $m_{f}=40\ eV$, $\rho_{0}=5\ GeV/cm^{3}$ . }\label{typical}
\end{figure*}  
We will model the dark matter content in galaxies as a core consisting of self-gravitating, self-interacting  degenerate fermions with an effective equation of state given by Eq. \eqref{polytrope}. 
We will consider the system to be spherically symmetric and static. The density profile of the core  is found by solving the Tolman-Oppenheimer-Volkoff (TOV) equations. Dark matter cores and halos have low compactness, that is, the system mass is small compared to the radius $GM/r\ll 1$ and $p \ll \rho$, which simplifies the TOV system of equations to the Newtonian TOV equations given by

\begin{equation} 
\label{tov}
    \frac{dp}{dr}=-\frac{GM\rho}{r^{2}}, \quad   \frac{dM}{dr}=4\pi r^{2}\rho.
\end{equation}

\begin{figure*}
    \centering
   \includegraphics[width=0.45\textwidth]{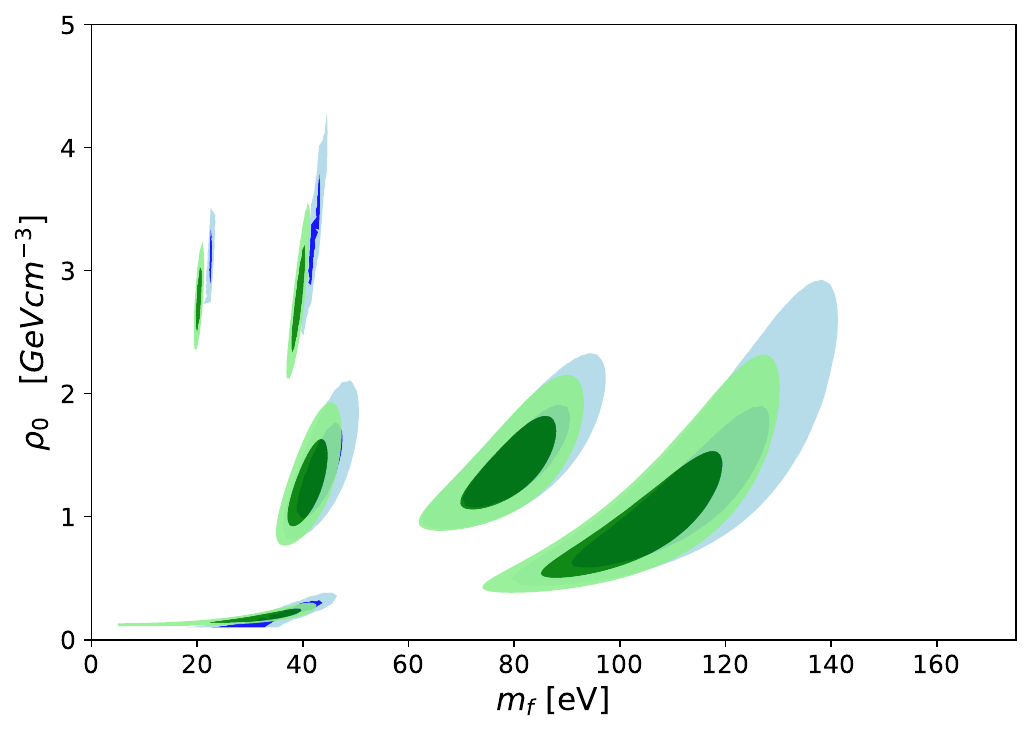}
\includegraphics[width=0.45\textwidth] {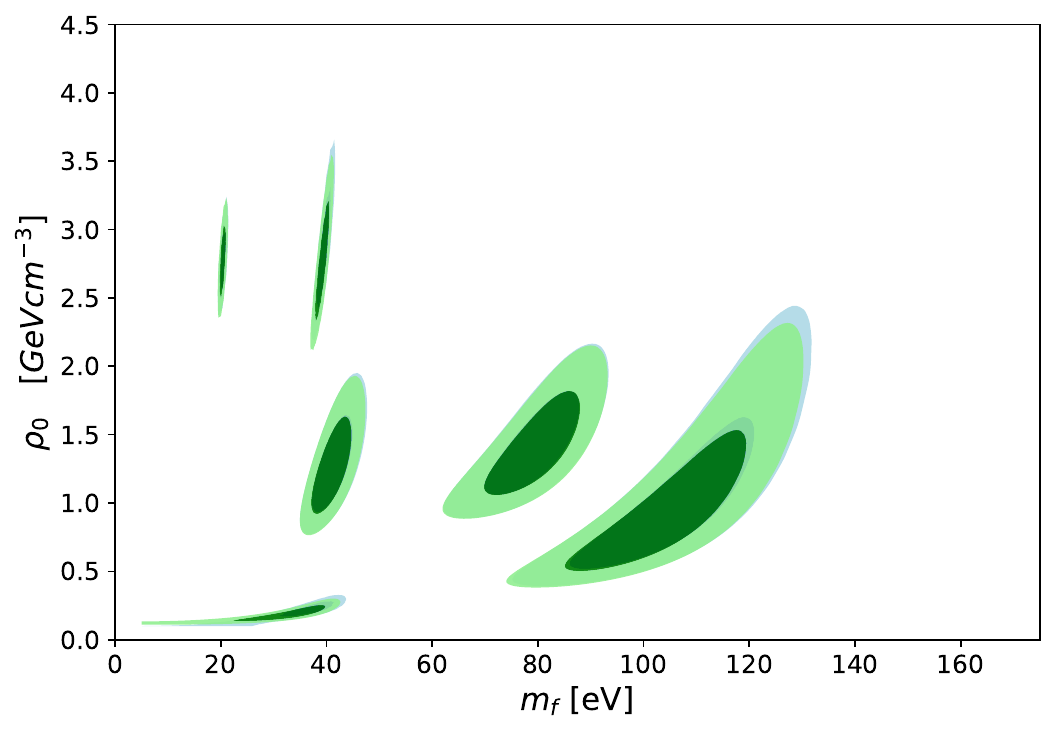}
    \caption{
Contours of the masses of the fermion and central densities allowed by the fit to the rotational curves of the 6 LSB galaxies studied, in the case of no self-interaction. The darker and lighter regions correspond to the 68\% and 95\% confidence limits, respectively. The fully degenerate model is shown in green, while the semi-degenerate fermion with a NFW halo (left panel) and with a perfect-fluid halo (right panel) are shown in blue.}\label{y=0 plot}
\end{figure*}

\begin{figure*}
    \centering
    \includegraphics[width=0.45\linewidth]{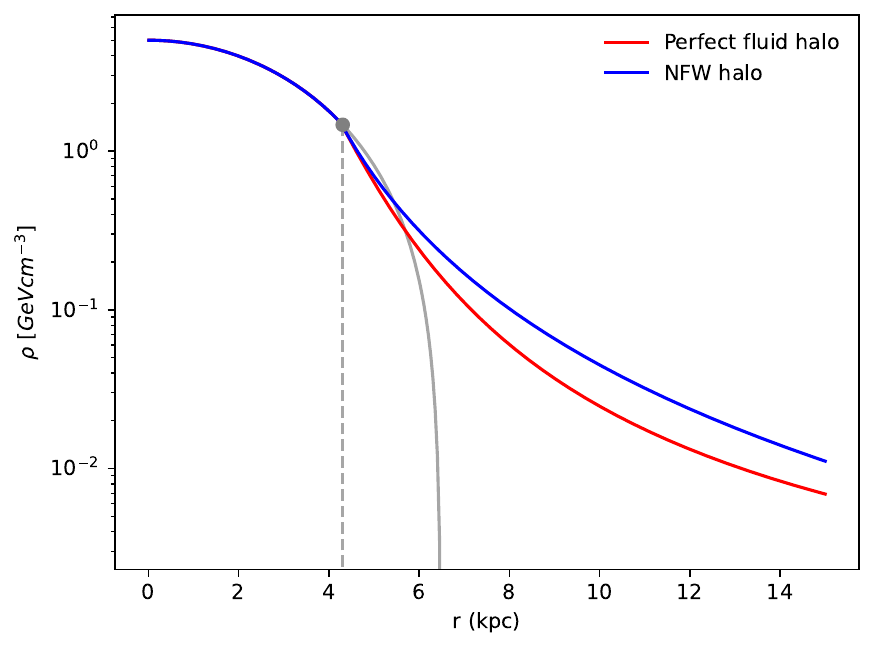}
    \includegraphics[width=0.45\linewidth]{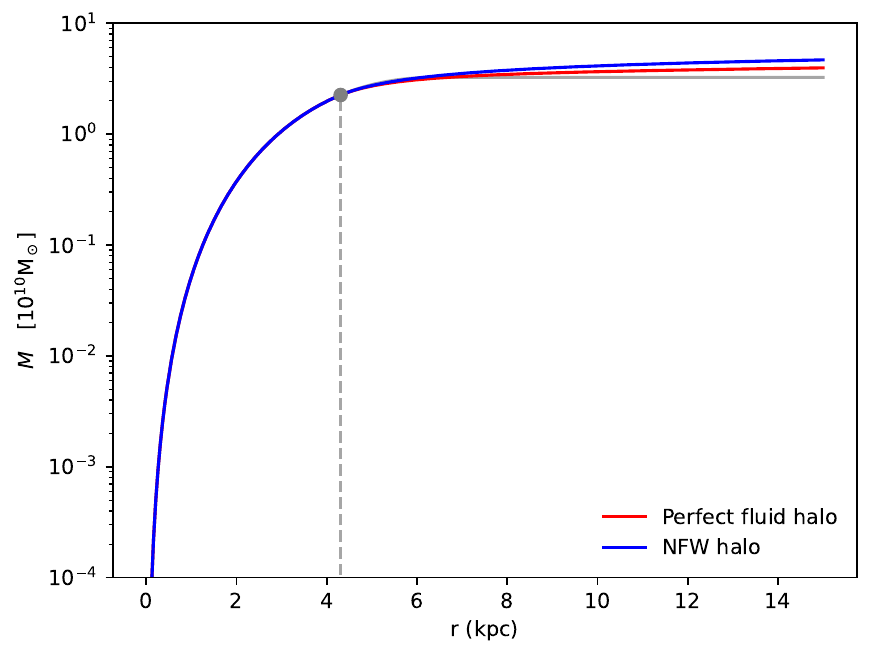}
    \caption{Density and mass profiles for the core-halo model with the transition point from a degenerate to a non-degenerate system. $\rho_0 =5 GeV/cm^{3}$, $m_{f}=50 eV$, $y=100$}
    \label{fig: density-profiles}
\end{figure*}

In the case of our equation of state \eqref{polytrope} the TOV system can be expressed as 

\begin{equation} 
\label{tov rho}
    \frac{d\rho}{dr}=-\left(\frac{3}{\pi^{4}}\right)^{1/3}\frac{GM}{r^{2}}m_{f}^{8/3}\rho^{1/3}\left[1+2y^{2}\left(\frac{3\rho}{\pi^{4}m_{f}^{4}}\right)^{1/3}\right]^{-1},
\end{equation}

\begin{equation}
\label{tov M}
    \frac{dM }{dr}=4\pi r^{2}\rho
\end{equation}

This TOV system is solved numerically, with boundary conditions $M(r=0)=0$ and $\rho(r=0)=\rho_{0}$ , where $\rho_{0}$ is the central density of the self-gravitating object and it is a free parameter. \\

The system's free parameters affect both the size and total mass of the halo. Specifically, lighter fermion masses lead to larger objects, while heavier masses result in smaller ones. Additionally, as the self-interaction parameter increases, the total mass of the system also rises.
In figure \ref{typical}, some numerical solutions to the TOV system (Eqs. \eqref{tov rho} and
\eqref{tov M}) for different elections on the free parameters $m_f$ and $y$ are presented. Left panel of Fig. \ref{typical} illustrates the effect of the mass of the fermion for fixed values of $\rho_0=5~\rm{GeV/cm}^3$ and $y=100$. On the other hand, right panel shows the changes in the mass and density profiles as a function of the self-interaction strength constant $y$ for fixed values of $m_f=40~\rm{eV}$ and $\rho_0=5~\rm{GeV/cm}^3$. Observe that the galactic scale of the self-gravitating object is mainly determined by the ultra light value of $m_f$.  
\\

This description is valid as long as the temperature of the system sits below the degeneracy temperature. The temperature of the system can be determined at any point using the virial theorem

\begin{equation}
    T(r)=\frac{2}{7}\frac{GMm_{f}}{K_{b}r},
\end{equation}

where $M$ is the enclosed mass of the compact object, and r the radial distance where the temperature is computed. Note that the factor $\frac{2}{7}$ comes from the fact that for a polytrope \cite{DKoester_1990}

\begin{equation}
    E_{grav}=-\frac{3}{5-n}\frac{GM^2}{r},
\end{equation}
and for a degenerate fermion $n=3/2$.\\

If the temperature of the core becomes high enough, according to the virial theorem, the conditions for a fully degenerate gas of fermions are not existent and it will transition into a halo, of which we have chosen two different options: a perfect fluid or a NFW type halo, as  shown visually in Fig. \ref{fig: density-profiles}. 

If the gas thermalized, then it would lead to the equation of state of a non-relativistic classical gas at statistical equilibrium: the classical isothermal equation of state $p=\omega\rho$.\\

We require hydrostatic equilibrium, and for that to be the case, we need that $p_{core}(r_{*})=p_{halo}((r_{*})$, where $r_{*}$ is the point where the gas stops being degenerate. This means that $\omega$ is related to $r_{*}$ and the model parameters with the equation

\begin{equation}
    \omega=\frac{(3\pi^{2})^{2/3}}{5}\frac{\rho_{*}^{2/3}}{m_{f}^{8/3}}+\frac{y^{2}\rho_{*}}{m_{f}^{4}}.
\end{equation}

After the system stops being degenerate the density profile will be obtained by numerical integration. Starting from $r_{*}$, the equation for the density of the TOV has the form of

\begin{equation} 
\label{tov pf}
    \frac{d\rho}{dr}=-\frac{1}{\omega}\frac{GM\rho}{r^{2}}.
\end{equation}

The other possibility to model the core-halo is the following: 
we have the same degenerate fermion gas as a core, but as a halo it will have the empirically obtained, and standard, density profile of the $\Lambda$CDM model: the NFW profile. The NFW density profile is defined as follows \cite{1997Navarro}

\begin{equation}
    \rho_{NFW}(r)=\frac{\rho_{s}}{\frac{r}{R_{s}}\left(1+\frac{r}{R_{s}}\right)^{2}},
\end{equation}

where $\rho_{s}$ is the scale density and $R_{s}$ is the scale radius. The system is in virial and dynamical equilibrium, that produce an effective pressure, but in the case of the NFW profile there is no pressure as $\omega=0$. To ensure a good transition from the degenerate fermion core to the NFW modeled region, we demand that the density and its radial derivative to be continuous in the transition radius $r_{*}$. This results in the scale radius $R_{s}$ and the scale density $\rho_s$ being given by the expressions \\

\begin{equation}
R_{s}=-r_{\star}\left[\frac{r_{\star}\rho'_{\star} + 3\rho_{\star}}{r_{\star}\rho'_{\star}+\rho_{\star}}\right],\quad \rho_{s}=-\frac{4(r_{\star}\rho'_{\star}\rho_{\star}^{3} +\rho_{\star}^{4})}{(r_{\star}\rho'_{\star} + 3\rho_{\star})^{3}},
\end{equation}

where $\rho_{*}$ and $\rho'_{*}$ are the density and it's derivative in $r_{\star}$. \\
Both models for the dark matter core-halo system have only three free parameters: the mass of the fermion $m_f$, its inter-particle interaction $y$ and the central density of the system $\rho_0$. This parameters influence the size and mass of the system, creating a variety of different possible structures.

 In Figure \ref{fig: density-profiles} the density and mass profiles for the core-halo models are shown, illustrating the transition point from a degenerate to a
non-degenerate system for a perfect fluid halo (red line) and the NFW halo (blue line), for some representative values for $m_f$ and $y$. Observe that these are smooth profiles in both cases.

With these two core-halo models, we are able to explore the possibility of the inner region of a DM halo being composed of a degenerate gas of self-interacting fermions, capturing what would be the quantum effects of the central region of the halo.

The option of the NFW halo type provides a reference point for the standard results of CDM, which suffers from the \textit{core-cusp problem}, meanwhile the perfect fluid halo corresponds to the non-degenerate case of a fermionic gas, which is a perfect fluid with equation of state $p=\omega\rho$.

\section{Fitting Low Surface Brightness galaxies} \label{lsb}

\begin{figure*}
     \begin{subfigure}[b]{0.45\textwidth}
         \centering
         \includegraphics[width=\textwidth]{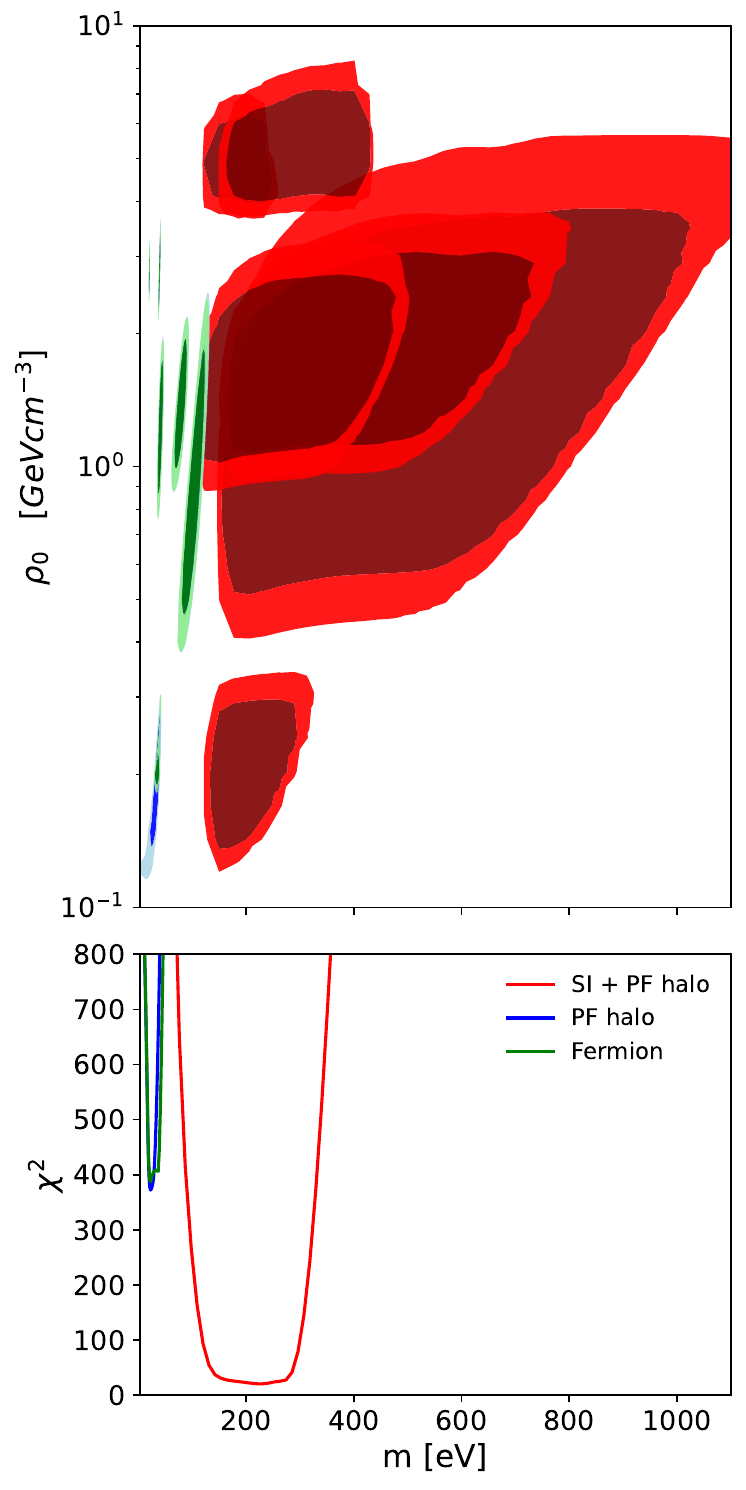}
         \caption{Perfect Fluid halo}
     \end{subfigure}
     \begin{subfigure}[b]{0.45\textwidth}
         \centering
         \includegraphics[width=\textwidth]{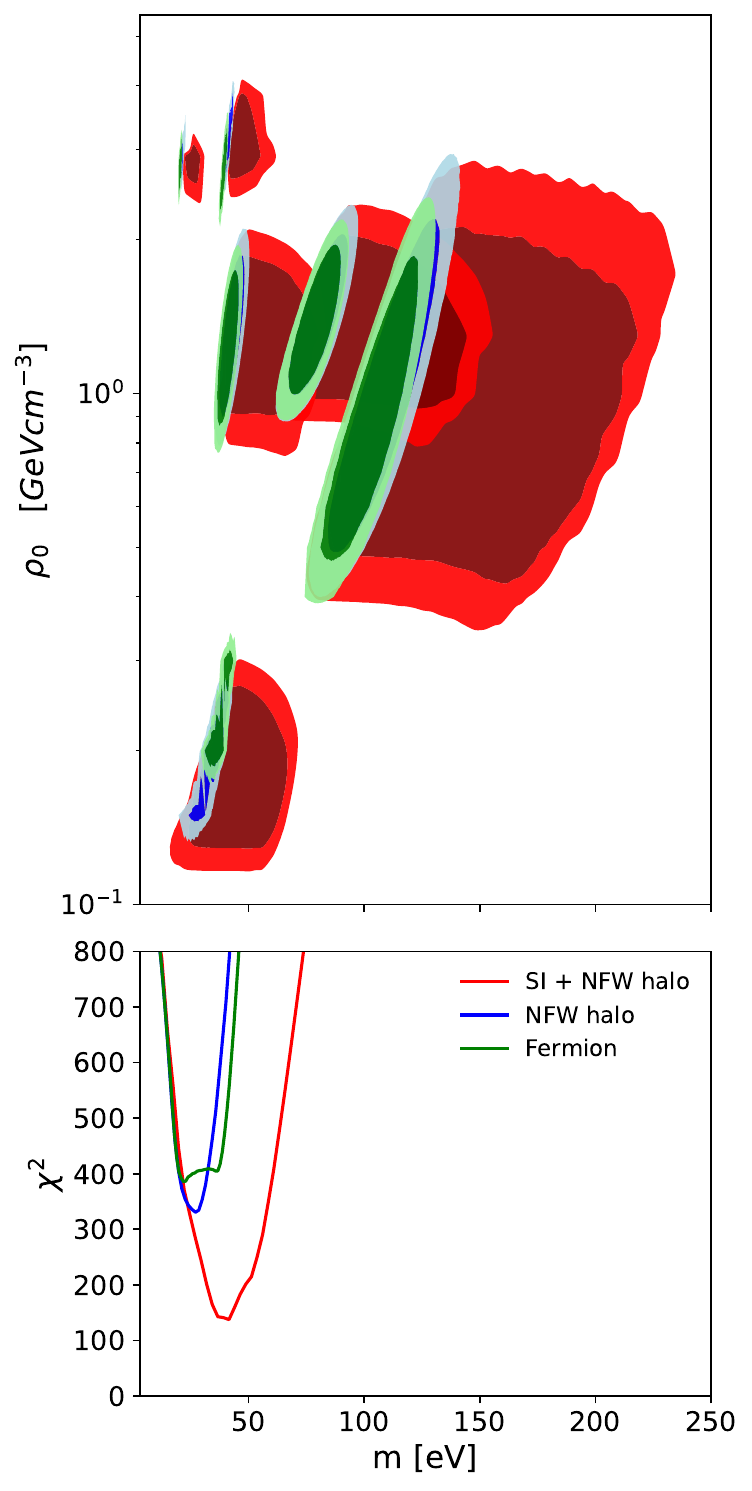}
         \caption{NFW halo}
     \end{subfigure}
        \caption{Comparison of fits for the perfect-fluid (a) and NFW (b) halo models. Top panels: contour for the 68\% and 95\%  confidence limits for the three possible cases: a self-interacting fermion with halo (red), a non-interacting fermion with halo (green), and the fully degenerate fermion (blue). Bottom panels: $\chi^{2}$ as a function of the fermion mass for each of the same models.
}\label{comparison-models}
\end{figure*}
Low Surface Brightness (LSB) galaxies are dominated by dark matter which make the inclusion of baryonic matter unnecessary and good systems to study dark matter properties in galactic scales \cite{deBlok1997}. For instance, LSB galaxies have shown that CDM is not an appropriate model to describe them \cite{deBlok2001}.  
In our case, the dark matter particle is a fermion with mass $m_f$ and self-interaction strength $y$ that fulfills equation of state Eq. \eqref{polytrope}. \\

\begin{table}
\caption{Characteristics of the LSB galaxies selected for analysis.} \label{galaxies info}
\begin{ruledtabular}
\begin{tabular}{l |cccc}
Galaxy & Q\footnote{H$\alpha$ rotation curve quality: 1 = good; 2 = fair; 3 = poor} & Type\footnote{Hubble type from \cite{deVancoleurs1991}, \cite{Schombert1992}  or the NASA/IPAC Extragalactic Database (NED)} & \begin{tabular}{c}
$r_{\max}$ \\
{$[kpc]$}
\end{tabular} & $N_{data}$ \\
\hline
ESO 0140040 & 1 & Sb & 29.2 & 8  \\
ESO 0840411 & 1 & Pec. & 8.9 & 9  \\
ESO 1200211 & 2 & Sa & 3.5 & 14 \\
ESO 1870510 & 1 & dSB & 3.04 & 11 \\
ESO 2060140 & 1 & Sc & 11.65 & 15 \\
ESO 3020120 & 1 & dS & 10.95 & 11 \\
\end{tabular}
\end{ruledtabular}
\end{table}

We chose a subset of six of the LSB galaxies presented in \cite{McGaugh2001}, which include galaxies with good or fair quality of data and different morphologies, as presented in Table \ref{galaxies info}. In order to obtain the parameters that best reproduce the rotational velocity curves of the six LSB galaxies selected we do a minimum $\chi^{2}$ analysis to obtain the best-fit values for the fermion mass $m_{f}$, the self-interaction parameter $y$, and the central density $\rho_0$.  

\begin{equation}
    \chi^{2}(\rho_{0},m_{f},y)=\sum_{i}\left(\frac{v^{th}(r,\rho_{0},m_{f},y)-v_{i}^{obs}}{\delta v_{i}^{obs}}\right)^{2},
\label{eq_chi}
\end{equation}

where $v_i^{obs}$ is the observed velocity, $\delta v_{i}^{obs}$ the error in the measurement of $v_i$ and $v^{th}$ is the theoretical value of the velocity, which is computed at the same position where $v_i^{obs}$ was measured. The theoretical value of the velocity is computed as 

\begin{equation}
\label{velocity}
v^{th}(r,\rho_0.m_f,y)=\sqrt{\frac{GM(r)}{r}},   
\end{equation}

where the value of $M(r)$ is the result of our self-gravitating system. If the system is below $T_{deg}$ it is obtained through solving numerically the TOV system, presented in equations \ref{tov rho} and \ref{tov M},  with a Runge-Kutta 4th order method and a step size of $h=0.05$ kpc, which entails an accustomed accumulated error of $O(h^4)$. If the temperature of the system surpasses the degeneracy threshold value, then we need to compute the corresponding NFW mass profile, or integrate numerically the TOV system, in the case of a perfect fluid halo, with the corresponding equation for the density, presented in equation \ref{tov pf}. The mass of the system is conformed exclusively by the dark matter halo, due to this being LSB galaxies.\\

To understand the role of the self-interactions and the influence of the different type of halos, we performed fits for three different cases: a purely degenerate fermion ($y=0$, without halo), a core–halo system (NFW or perfect fluid, $y=0$), and a core–halo with self-interaction ($y\neq0$). 

\subsection{No Self-interaction ($y=0$)}

Similar models for the dark matter halo considered degenerate fermions with no self-interaction
\cite{Randall2017,Destri2013,Destri2013a,Domcke2015,barranco2019constraining,deVega:2013ysa}. 
In the present work, the same model for the dark matter halo is assumed and a $\chi^{2}$ analysis for a fully degenerate fermion dark matter halo with no interaction is done, i.e. $\chi^2=\chi^2(\rho_0,m_f,y=0)$. The constraints for the mass of the fermion and the central density (since we are fixing $y=0$) are shown in figure \ref{y=0 plot} in darker ($68\%$ C.L.) and lighter ($95\%$ C.L) green.\\

In contrast to the models in \cite{Destri2013,Destri2013a,Domcke2015,barranco2019constraining,deVega:2013ysa}, in our case, we consider two models for the halo. By repeating the same analysis for the LSB galaxies but with the inclusion of the NFW halo or the perfect fluid halo, we obtain new allowed regions for $m_f$ and $\rho_0$.  They are shown in Fig. \ref{y=0 plot} in darker ($68\%$ C.L.) and lighter ($95\%$ C.L) blue.\\

In both cases we may conclude that the mass of the fermion are quite narrow and differ from galaxy to galaxy, needing a different value for the fermion mass to fit each of the LSB galaxies studied. Adding a halo to the degenerate fermion core doesn't improve substantially, but it does broaden the allowed regions of the $m_f$ parameter. Specially for the case where the halo is modeled as a perfect fluid (see left panel of Fig. \ref{y=0 plot}). 

\subsection{With Self-interaction $y \ne 0$}

\begin{table*}
\caption{Best fit values for a core-halo system with a core of a degenerate fermion with self-interaction and a NFW halo of perfect fluid halo.} \label{SI best fit}
\begin{ruledtabular}
\begin{tabular}{l |cccc |cccc}
& \multicolumn{4}{c|}{Perfect Fluid Halo} & \multicolumn{4}{c}{NFW Halo} \\
Galaxy & $\chi^{2}/\mathrm{dof}$ & \begin{tabular}{c}
$m_f$ \\
$[eV]$
\end{tabular} & $y$ & \begin{tabular}{c}
$\rho_0$ \\
$[\mathrm{GeV/cm^3}]$
\end{tabular} & $\chi^{2}/\mathrm{dof}$ & \begin{tabular}{c}
$m_f$ \\
$[eV]$
\end{tabular} & $y$ & \begin{tabular}{c}
$\rho_0$ \\
$[\mathrm{GeV/cm^3}]$
\end{tabular} \\
\hline
ESO0140040 & 0.982 & 198.485 & 3684.031 & 4.771 & 3.646 & 25.882 & 64.037 & 2.902 \\
ESO0840411 & 0.062 & 196.566 & 5139.043 & 0.193 & 0.099 & 35.941 & 52.160 & 0.192 \\
ESO1200211 & 0.080 & 270.707 & 1239.429 & 1.160 & 0.184 & 108.824 & 57.794 & 0.944 \\
ESO1870510 & 0.045 & 428.788 & 4417.470 & 1.721 & 0.193 & 85.294 & 96.519 & 1.302 \\
ESO2060140 & 0.353 & 463.636 & 8091.108 & 5.487 & 2.257 & 41.765 & 28.187 & 2.944 \\
ESO3020120 & 0.003 & 190.909 & 1893.240 & 1.605 & 0.184 & 44.118 & 38.344 & 1.302 \\
\end{tabular}
\end{ruledtabular}
\end{table*}

\begin{figure}[h!]
    \centering
    \includegraphics[width=0.5\textwidth]{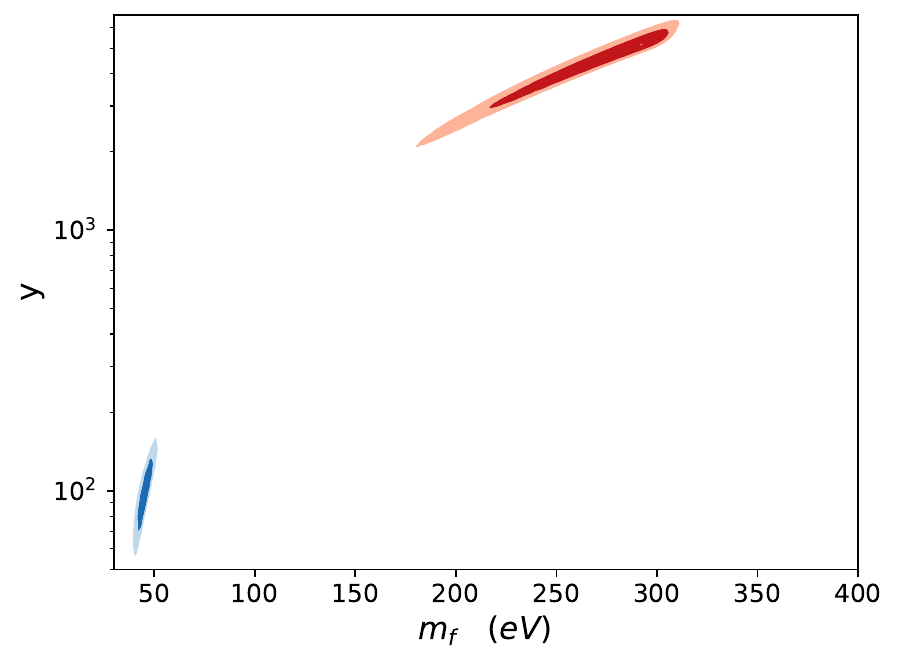}
    \caption{Masses and interaction parameters allowed to fit the rotational curve data of the different galaxies studied. The contours represent the 68\% (darker shade) and 95\% (lighter shade) confidence limits. In red we present the contours for the case with a perfect fluid halo, in blue the NFW halo case.}
    \label{combined chi self interaction}
\end{figure}

We arrive at the central point of this work: can self-interaction alleviate the need of different fermions for each galaxy under study?
In order to answer this question, we perform a $\chi^2$ analysis with $y\ne0$, i.e. $\chi^2=\chi^2(\rho_0,m_f,y \ne 0)$. The results for the projected values allowed for $m_f$ and $\rho_0$ 
are shown in Fig.\ref{comparison-models}. Color code for $y \ne 0$ is: darker red isocurves for $68\%$ C.L. and lighter red isocurves for $95\%$ C.L.  For comparison we have included in the same plot as in the previous section for the case with no self-interaction, $y=0$. For $y=0$ isocurves, the color code is the same as in Figure \ref{y=0 plot}.\\

It can be seen that the core–halo model broadens the allowed parameter space, and these regions become even larger when self-interactions are included. The figure \ref{comparison-models} also display the projection of the sum of the $\chi^{2}$ as a function of the fermion mass, $m_f$, the parameter that is required to be common to all galaxies. The $\chi^2$ curves indicate that the case with both a halo and self-interactions provides fits that are comparable to, or better than, those obtained with the previous cases. \\

The fit of the rotational curves was performed by a $\chi^2$ minimization over a three-dimensional parameter grid to obtain the best-fit values for the fermion mass $m_{f}$, the self-interaction parameter $y$, and the central density $\rho_0$. The grid resolution was  $100\times100\times100$ ($10^6$ points). The self-interaction parameter was sampled using a log-uniform (geometric) spacing, while the remaining parameters were sampled linearly within their respective ranges.\\

The best fit points obtained are presented in Table \ref{SI best fit}. They show that $\chi^2/\rm{dof}$ is smaller than unity for the halo modeled as a ideal gas, while the halo modeled with a NFW density profile gives reasonable fits with $\chi^2/\rm{dof} \sim 1$. This results suggest that a reduction on the free parameters is possible. In particular, we are interested in the dark matter particle properties, which for our case are $m_f$ and $y$. To obtain the information of the two parameters that are of interest, the fermion mass and the self interacting parameter, we did a combined minimum $\chi^2$ :  $\chi^2_{combined} = \sum_{i=1}^{n} \chi^2_{p_i}$,  where $\chi^2_{p_i}$ is the profiled chi-squared statistic for the $i$-th dataset. \\

The results of the combined fit minimized in $\rho_0$ and projected in the $(m_f,y)$ plane are shown in Fig. \ref{combined chi self interaction} where we present the contours of the parameters that fit the data with 1$\sigma$ (68\% C.L.) and 95\% confidence levels. Dark and light blue are isocurves for the degenerate core with a NFW halo while the dark and light red correspond to isocurves for the degenerate core with a ideal gas equation of state for the halo. 
In summary, the fit values of the combined $\chi^2$ are: 
\begin{itemize}
\item For the NFW halo 
\begin{equation}
m_f=44.62^{+4.61}_{-2.31}~\rm{eV}, \ y=95.92^{+36.73}_{-24.49}\,,\label{bfp_nfw}
\end{equation}
\item For the perfect fluid halo 
\begin{equation}
m_f=(2.93^{+0.2}_{-1.38})\times 10^{2}~\rm{eV}, \ y=(5.07^{+0.93}_{-3.61})\times 10^{3}\,.\label{bfp_pf}
\end{equation}
\end{itemize}.
\\
There is only one free parameter left in the model: the central density of the system, a value which is case specific for each of the galaxies. 
We finish this section by fixing $m_f$ and $y$ in the equation of state Eq. \eqref{polytrope} and perform a $\chi^2=\chi^2(\rho_0)$ analysis of our sample rotational data of the six LSB galaxies. These fits gave good results and are presented in Tables \ref{perfect fluid fixed} for the model with a halo with a barotropic equation of state  and \ref{NFW fixed} for a halo with a NFW density profile.  \\

\begin{figure*} 
    \includegraphics[width=\textwidth]{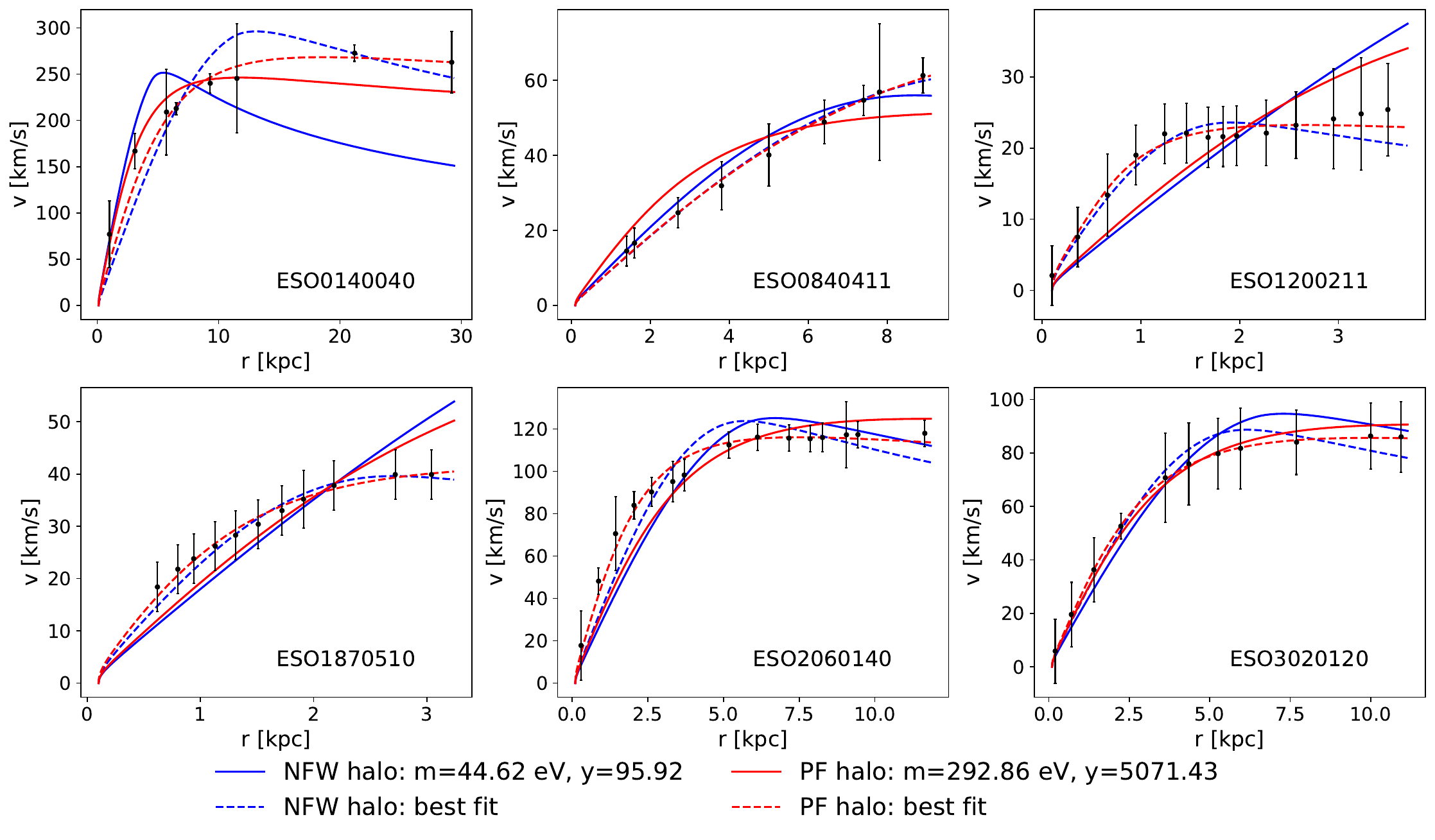}
    \caption{Rotation curves for the LSB galaxy sample. Solid curves show the best-fit results, shown in Table \ref{SI best fit}; dashed curves  show the predicted curves when the fermion mass mf and self-interaction parameter y are fixed. Perfect-fluid halos (red) use $m_f$=292.86eV, y=5071.43; NFW halos (blue) use $m_f$=44.62eV, y=95.92. The central density values for each of the galaxies and $\chi^{2}/dof$ are presented in Tables \ref{perfect fluid fixed} and \ref{NFW fixed} }
    \label{LSB_rot_curves}
\end{figure*}

In general, even with one free parameter, a self-interacting fermion with a defined mass and self-interaction strength fixed  gives reasonable fits. For reference, the predicted rotational curve and a comparison with the data  are shown in Fig. \ref{LSB_rot_curves}, where we also have plotted the predicted rotational curve for the best fit values obtained by minimization of $\chi^2$ by varying the three parameters as shown in Table \ref{SI best fit}.

\begin{table}[h]
\caption{Best fit results for a model with a perfect fluid halo and fixed parameters $m_f=292.86eV$ and $y=5071.43$.} \label{perfect fluid fixed}
\begin{ruledtabular}
\centering
\begin{tabular}{lcc}
Galaxy & \begin{tabular}{c}
$\rho_0$ \\
$[\mathrm{GeV/cm^3}]$
\end{tabular} & $\chi^{2}/\mathrm{dof}$ \\
\hline
ESO0140040 & 10.028 & 3.558 \\
ESO0840411 & 0.442 & 1.886 \\
ESO1200211 & 0.329 & 0.999 \\
ESO1870510 & 0.835 & 1.128 \\
ESO2060140 & 2.580 & 2.171 \\
ESO3020120 & 1.361 & 0.081 \\
\end{tabular}
\end{ruledtabular}
\end{table}

\begin{table}[h]
\caption{Best fit results for a model with a NFW and fixed parameters $m_f=44.62eV$ and $y=95.92$.}
\label{NFW fixed}
\begin{ruledtabular}
\centering
\begin{tabular}{lcc}
Galaxy & \begin{tabular}{c}
$\rho_0$ \\
$[\mathrm{GeV/cm^3}]$
\end{tabular} & $\chi^{2}/\mathrm{dof}$ \\
\hline
ESO0140040 & 11.140 & 24.904 \\
ESO0840411 & 0.249 & 0.392 \\
ESO1200211 & 0.268 & 1.427 \\
ESO1870510 & 0.715 & 1.808 \\
ESO2060140 & 1.983 & 3.177 \\
ESO3020120 & 0.976 & 0.428 \\
\end{tabular}
\end{ruledtabular}
\end{table}

\section{The Milky Way Rotational Curve} \label{sec:mw}
\begin{figure*}
    \centering
\includegraphics[width=0.95\textwidth]{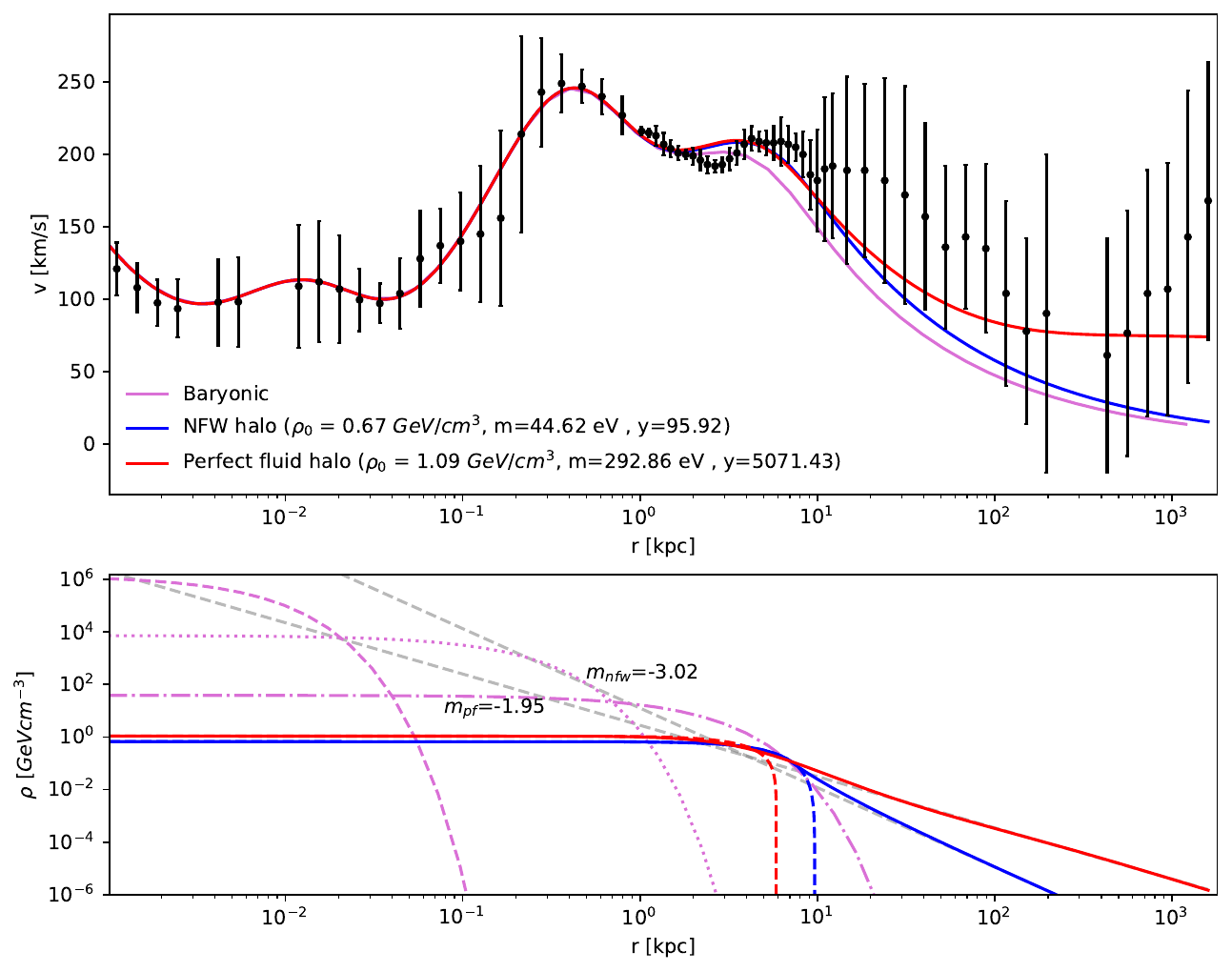}
    \caption{Milky Way rotational velocity curve obtained for the two core-halo models with fixed values of the fermion mass and the self-interaction parameter given by Eqs. \eqref{bfp_nfw} and \eqref{bfp_pf}. Plots were done with the best fit for the only free parameter $\rho_0$. For the NFW-type halo, the fit return a bets fit for the central density of $\rho_0=0.67\ \rm{GeVcm}^{-3}$, whereas the perfect-fluid halo requires $\rho_0=1.09\ \rm{GeVcm}^{-3}$. The lower panel shows the corresponding density profiles for the different components of the Milky Way.}
    \label{fig:MWfit}
\end{figure*}
Having found values of the fermion mass and interaction parameter that reproduce well enough the rotational curves of the LSB galaxies (see Eqs. \eqref{bfp_nfw},\eqref{bfp_pf}), we can use these values to predict the rotational curve of different systems. First, we will test the model with the Milky Way's rotational curve. 
 Measurements of the rotational velocity of the Milky Way are divided in components related to the structure of the galaxy: the central black hole, the inner bulge or core of the galaxy, the main bulge and the disk of the galaxy. In addition to these four elements, we will add a dark matter core-halo as we have explained in previous section. \\

The data 
covers a wide range of radius: from 1 pc to several hundred kpc, without a gap of data points. This was achieved by combining data from the outer disk  and analyzing high-resolution longitude-velocity diagrams of the galactic center, observed with the Nobeyama  telescope in the CO and CS line emissions. There are 67 data points which go from 1.12 pc to 1,600 kpc \cite{2009Sofue,Sofue2013}. \\

The disk, inner bulge and main bulge can be modeled by an exponential sphere model, in which the mass density of the region $\rho_{R}(r)$ is given by:
\begin{equation}
    \rho_{R}(r)=\rho^{c}_{R}\exp(-r/a_{R}), \label{spheremodel}
\end{equation}

where $\rho^{c}_{R}$ corresponds to the central density of the region $R$, and $a_{R}$ to its scale radius. 
The values of the parameters $\rho^{c}_{R}$ and $a_{R}$ are fixed to the values shown in table \ref{valores_} \cite{2009Sofue,Sofue2013}.

Integrating the Eq. \eqref{spheremodel} for the mass density we are able to find the mass as a function of the radius $M_{R}$

\begin{equation}
\label{mass eq}
    M_{R}(r)=8\pi a^{3}_{R}\rho^{c}_{R}\left(1-e^{-r/a_{R}}\left(1+\frac{r}{a_{R}}+\frac{1}{2}\left(\frac{r}{a_{R}}\right)^{2}\right)\right).
\end{equation}

The velocity of a test particle subject to the mass $M_R(r)$ will be given by
\begin{equation}
    v_R^2(r)=\frac{G M_R(r)}{r}\,,\label{vel_r}
\end{equation}
with $R=IB,MB,D$. 
Saggitarius A$^*$, the supermassive black hole located at the center of the Milky Way contributes to the total mass, hence, to the velocity of the stars in the galaxy. This contribution will be modeled as a point mass and given by
\begin{equation}
    v_{BH}^2=\frac{GM_{BH}}{r}\,.
\end{equation}

 The mass of the central black hole  has been constrained to $4.1\pm0.6 \times 10^{6}M_{\odot}$ \cite{Ghez:2008ms}. We used the central mass value to calculate the rotational velocity contribution of the Saggitarius A$^*$.

 $v_{DM}(r)$ is computed with Eq. \eqref{velocity} with the mass of the dark matter halo obtained by solving equations \eqref{tov} for the two models for the halo.\\
\begin{table}[h!]
\caption{Values used to fit the different sections of the Milky Way rotational velocity \cite{2009Sofue,Sofue2013}.}\label{sofue}
\begin{ruledtabular}
\begin{center}
\begin{tabular}{lcc}
Region (R) & \begin{tabular}{c}
Total Mass\\ $[10^{10}M_{\odot}]$\end{tabular} & \begin{tabular}{c}Scale Radius $a_{R}$\\ $[kpc]$\end{tabular}\\ 
\hline
Inner Bulge (IB)                         & $5.0 \times10^{-3}$ & $3.8 \times10^{-3}$  \\
Main Bulge (MB)  & $8.4 \times10^{-1}$ & $1.2 \times 10^{-1}$  \\
Disk (D)  & 4.4  & 1.3   \\ 
\end{tabular}
\end{center}
\label{valores_}
\end{ruledtabular}
\end{table}
The theoretical rotational velocity can be obtained from the expression
\begin{equation}
    v^{th}(r)=\sqrt{\sum_{R=IB,MB,D} v^{2}_{R}(r) + v^{2}_{BH}(r)+v^{2}_{DM}(r) },
    \label{vel_rot}
\end{equation}
where $IB$ is the inner bulge, $MB$ is the main bulge, $D$ is the disk, $BH$ corresponds to the black hole and $DM$ to the dark matter halo. The velocity from each section of the Milky Way is equation \ref{vel_r}, where the mass of IB, MB and D is obtained from equation \eqref{mass eq}.

Having fixed the scale radius $a_R$ and central densities of the baryonic components $\rho_R^c$ of the Milky Way with the values reported in Table \ref{sofue}, then the baryonic content of the Milky Way is fixed.  By fixing the fermion mass  $m_f$ and the self interaction parameter $y$ with the central values of Eqs. \eqref{bfp_nfw} and \eqref{bfp_pf} for the NFW and perfect fluid halos, the only free parameter in $v^{th(r)}$ to fit the rotational velocity data of the Milky Way is the central density of the core-halo system that model the dark matter halo. The theoretical rotational velocity obtained from computing \ref{vel_rot} was compared and fitted to the values reported in \cite{2009Sofue,Sofue2013}. The best fit point for the rotational curve of the Milky Way is $\rho_0=0.67~\rm{ GeV/cm}^{3}$  with a good fit given by a $\chi^{2}/dof=0.93$ for the case with a NFW halo.
Similarly, we get $\rho_0=1.09~\rm{ GeV/cm}^{3}$ with a  $\chi^{2}/dof=0.88$ for the case with a perfect fluid halo.
In figure \ref{fig:MWfit} we plot the predicted rotational velocity curve for the two cases overlaped with the data reported by \cite{2009Sofue,Sofue2013}. The blue line corresponds to the case of a NFW halo while the red line is the case for a perfect fluid with barotropic equation of state. For comparison we have included the case where no dark matter is included in a light purple line. In the bottom panel of Fig. \ref{fig:MWfit} is plotted the density profile, including the baryonic contribution. 
Baryonic matter dominates the stellar dynamics at radius smaller than one kilo-parsec, and dark matter stars dominate only for larger radius. The halo with a barotropic equation of state fits better that the NFW halo. For $r \gg 1~$kpc the halo mass density $\rho(r)$ follows a power law decay. In particular, as expected, the NFW decays as $\rho(r)\sim r^{-3}$ while the perfect fluid halo decays as $\rho(r)\sim r^{-2}$ as can be seen in the lower pannel of Fig. \ref{fig:MWfit}. This change in the decay of the mass density implies that the total mass of the core-halo system differs. These differences could be important in the determination of the stellar dynamics and thus lead to possible routes towards a better understanding of the structure of the dark matter halos.  \\

\section{Milky Way's Dwarf galaxies}

Now we test both core-halo models with the {\it classical} Milky Way's dwarf spheroidal galaxies. Those galaxies are low-luminosity, low-surface-brightness, satellite galaxies characterized by a high-quality data sets of the projected velocity dispersion along the line-of-sight as a function of the projected radius $\sigma_{LOS}(r)$.
In these galaxies, rotation is negligible and the stars are in equilibrium against gravity by their random motion. 
Assuming spherical symmetry, with constant orbital anisotropy $\beta$, the observed projection of the velocity dispersion along the line-of-sight at the projected radius $r$ is given by
\begin{equation}
\sigma_{LOS}^2(r)=\frac{2G}{I(r)}\int_{r}^{\infty}\nu(r')M(r')(r')^{2\beta-2}F(\beta,r,r')dr'\,.\label{sigmalos}
\end{equation}
Here
\begin{equation}
\nu(r)=\frac{3L}{4\pi r^3_{half}}\frac{1}{(1+(r/r_{half})^2)^{5/2}}\,,
\end{equation}
is the 3-dimensional spherically symmetric stellar density with total luminosity $L$ for a Plummer profile. $r_{half}$ measures the radius within $50\%$  of a galaxy's total emitted light (luminosity) is contained.
Furthermore, for a Plummer profile for stellar density 
\begin{equation}
    I(r)=\frac{L}{\pi r_{half}^2}\frac{1}{(1+(r/r_{half})^2)^2}\,.
\end{equation}
The factor that parameterizes the projection over the sigh-of-line  is given by
\begin{equation}
    F(\beta,R,r')=\int_{R}^{r'}\left(1-\beta\frac{R}{r^2}\right)\frac{r^{-2\beta+1}}{\sqrt{r^2-R^2}}dr\,.
\end{equation}

We compute the theoretical $\sigma_{Los}$ where the information of the dark matter halo is encoded in the mass $M(r)$ of the halo  obtained by solving equations \eqref{tov} for the two models for the halo. 
For simplicity we have assumed $\beta=0$ which means that velocities are isotropic. Relaxing this assumption will just improve the fit. 
Empirical velocity dispersion profiles for the eight "classical" dwarf-spheroidal galaxies where obtained from  \cite{2009ApJ...704.1274W,Walker:2007ju,Mateo:2007xh,Walker_2009}.

We performed a $\chi^2$ analysis by computing the theoretical $\sigma_{LOS}(r)$ by means of Eq. \eqref{sigmalos} by fixing the mass of the dark matter fermion $m_f$ and the self-interaction strength $y$ for the two core-halo models. The only free parameter is the central density $\rho_0$. The results are shown in Tables
\ref{dwarf-fluid} and \ref{dwarf-nfw} for both the fluid and NFW halo models. As illustration, in Figure \ref{fig:dwarf} are shown the best fit for the dwarf spheroidal galaxy Leo I. As before, the model fits the data of dwarf spheroidal with only one free parameter left: the central density.

\begin{table}[h]
\caption{Best fit results for the dispersion velocities of the Milky Way satellite galaxies for a model with a perfect fluid halo and fixed parameters $m_f=292.86eV$ and $y=5071.43$.} \label{dwarf-fluid}
\begin{ruledtabular}
\centering
\begin{tabular}{lccc}
Galaxy & \begin{tabular}{c}
$\rho_0$ \\
$[\mathrm{GeV/cm^3}]$
\end{tabular}  &  $\chi^{2}$  &  $\chi^{2}/\mathrm{dof}$ \\
\hline
Leo I &  0.554  &  25.352 & 1.690\\
Leo II & 0.450  & 24.690 &2.469\\
Draco & 0.581 & 13.279 & 0.830\\
Carina & 0.450 & 123.857 & 4.763\\
Fornax & 0.456  & 213.908 & 4.551\\
Sculptor & 0.559 & 123.513 & 3.529\\
Sextants & 0.449  & 103.554 & 7.397\\
Ursa Minor & 0.613  & 37.277 & 2.662\\
\end{tabular}
\end{ruledtabular}
\end{table}

\begin{table}[h]
\caption{Best fit results for the dispersion velocities of the Milky Way satellite galaxies for a model with a NFW  halo and fixed parameters $m_f=44.62eV$ and $y=95.92$.} \label{dwarf-nfw}
\begin{ruledtabular}
\centering
\begin{tabular}{lccc}
Galaxy & \begin{tabular}{c}
$\rho_0$ \\
$[\mathrm{GeV/cm^3}]$
\end{tabular}  &  $\chi^{2}$  &  $\chi^{2}/\mathrm{dof}$ \\
\hline
Leo I  & 0.485 & 30.0627 & 2.004 \\
Leo II & 0.346 & 24.567 & 2.457 \\
Draco & 0.505 & 15.511 & 0.969 \\
Carina & 0.247 & 66.957 & 2.575 \\
Fornax & 0.366 & 260.274 & 5.538 \\
Sculptor & 0.505 & 140.991 &  4.028 \\
Sextants & 0.208 & 50.795 & 3.628 \\
Ursa Minor & 0.545 & 40.769 & 2.912 \\
\end{tabular}
\end{ruledtabular}
\end{table}

\begin{figure}[h]
    \centering
    \includegraphics[width=\linewidth]{dwarf_fit.eps}
    \caption{Fit for $\sigma_{LOS}$ for the dwarf spheroidal galaxy Leo I data. Upper panel corresponds to the perfect fluid halo, dashed lines are $1-\sigma$ deviation for the best fit parameter as reported in Table \ref{dwarf-fluid}. Lower panel correspionds to the NFW halo model.}
    \label{fig:dwarf}
\end{figure}

\section{Fitting the SPARC Database}\label{sec:sparc}
Next, we test the core-halo models with fixed $m_f$ and $y$ with a larger database. For definitiveness, we use  the \textit{Spitzer} Photometry and Accurate Rotation Curves (SPARC) database. It contains 175 galaxies that cover a wide spectrum of morphologies, luminosities and surface brightness \cite{Lelli_2016}. The database include both the data of the rotation curve of the galaxy and the inferred contribution of the baryonic components of the galaxy: the gas, the bulk and the disk. 
The theoretical velocity of rotation for a test particle is given by:

\begin{equation}
    V_{bar}=\sqrt{|V_{gas}|V_{gas} + \Upsilon_{disk} |V_{disk}|V_{disk} + \Upsilon_{bul} |V_{bul}|V_{bul} }
\end{equation}

where $\Upsilon_{disk}$ and $\Upsilon_{bul}$ are the stellar mass-to-light ratios of disk and bulge components, respectively \cite{Lelli_2016}.\\

Unlike LSB galaxies  where dark matter dominates the dynamics of the rotational curve, the baryonic matter present in the galaxy represent a significant part of what gives the rotational curve its shape. It is of special interest the mass-to-light ratio of the disk and bulge components of the galaxy, as modeling galaxy masses is often affected by uncertainties in the stellar mass-to-light ratio, which influence the relative contributions of the dark matter halo and the stellar disk to the rotational curve \cite{vanAlbada_1985}. To prevent this uncertainties and focus on the parameters of our model, we fix $\Upsilon_{disk}$ to 0.5 and $\Upsilon_{bul}$ to 0.7, these values have been estimated as acceptable for most galaxies at $\mu_{3.6}$ 
\cite{Eskew_2012, Oh_2008, Meidt_2014, Schombert_2018, Schombert_McGaugh_2014, McGaugh_2016}.\\

Taking this into consideration, we fitted the SPARC database with the values that best reproduce the six LSB galaxies analyzed in section \ref{lsb}, i.e. Eqs. \eqref{bfp_nfw} and \eqref{bfp_pf}. The $\chi^2$ function is computed in this case by computing the theoretical velocity 
\begin{equation}
    v^{th}=\sqrt{V_{bar}^2+v_{DM}^2(r)}\,,\label{vsparc}
\end{equation}
with $v_{DM}(r)$ computed as in the previous section \ref{sec:mw}. 

This procedure gave good results with a median $\chi^2/dof$ of 2.93 for the perfect fluid halo, meanwhile a 5.64 median value for the NFW case
(see figure \ref{fig:histogram}). The full results are presented in the Table \ref{table_sparc} in the appendix. In this table, the first column refers to the name of the galaxy.  The two cases for the halo model are reported: the NFW halo and the perfect fluid halo. For each case we report: the first column is the best fit value for the central density $\rho_0$ in units of GeV/cm$^3$ obtained form the fit as explained in Section \ref{sec:mw}. The second column shows the minimum value of the $\chi¨2$ function and the third column the f¡goodness of the fit expressed with the value of the minimum of the $\chi^2$ function divided by the degree of freedom.

A summary of the goodness of the fit for the SPARC database is represented in the histograms that are shown on the figure \ref{fig:histogram}. If the halo consists of a perfect fluid halo 58 galaxies (33\%) fall below $\chi^{2}/dof=1$, and 105 (60\%) fall below $\chi^{2}/dof=5$. On the other hand, in the case of a NFW density profile halo, 35 galaxies have $\chi^2/d.o.f$ lower than unity (20\%), and 83 galaxies with $\chi^2/d.o.f <5$ (47.43\%). 
Note that the only free parameter is the central density, thus a single fermion with mass of $m_f= 292.86~\rm{eV}$ ($m_f= 44.62~\rm{eV}$) and interaction strength constant $y=5.07\times10^3$ ($y=95.92$) can model nearly 200 cores of very different galaxies with a perfect fluid (NFW) halo.\\

\begin{figure}[h]
    \centering
    \includegraphics[width=\linewidth]{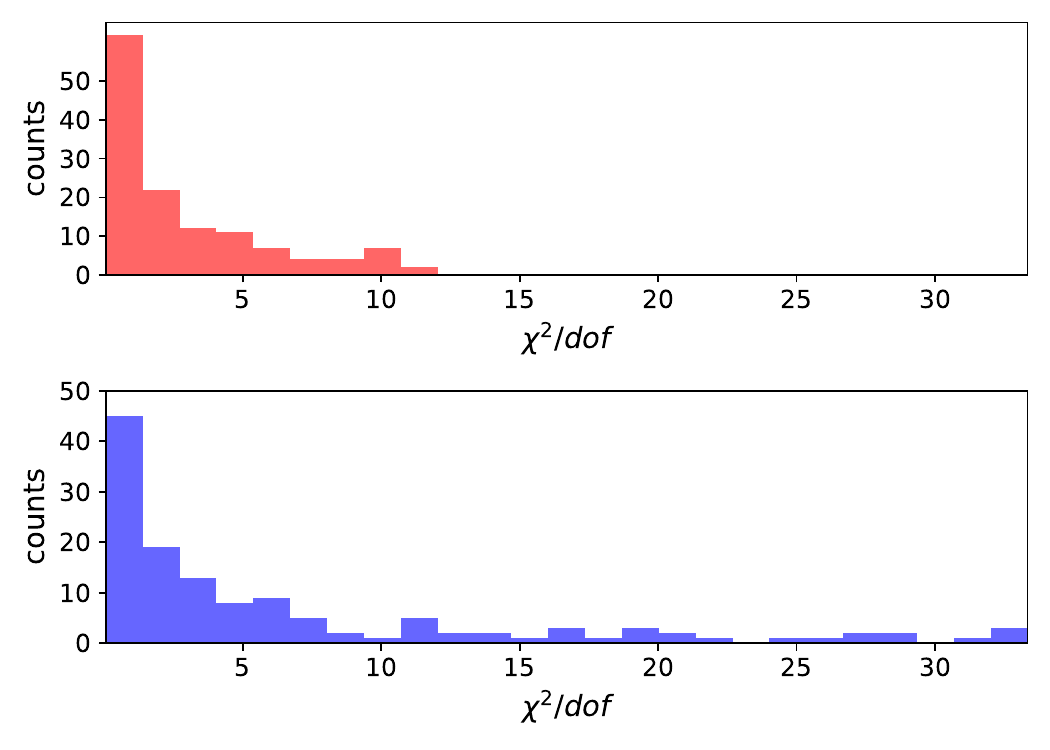}
    \caption{Distribution of $\chi^{2}/dof$ values obtained from the fits to the SPARC rotation-curve dataset. The upper panel shows the histogram for the perfect fluid case, while the lower panel displays the NFW halo case. In both cases the histogram is truncated at the 75th percentile in order to highlight the bulk of well-fitted galaxies.}
    \label{fig:histogram}
\end{figure}

\section{Discussion}\label{discussion}
\begin{figure*}
    \centering
    \includegraphics[width=0.95\textwidth]{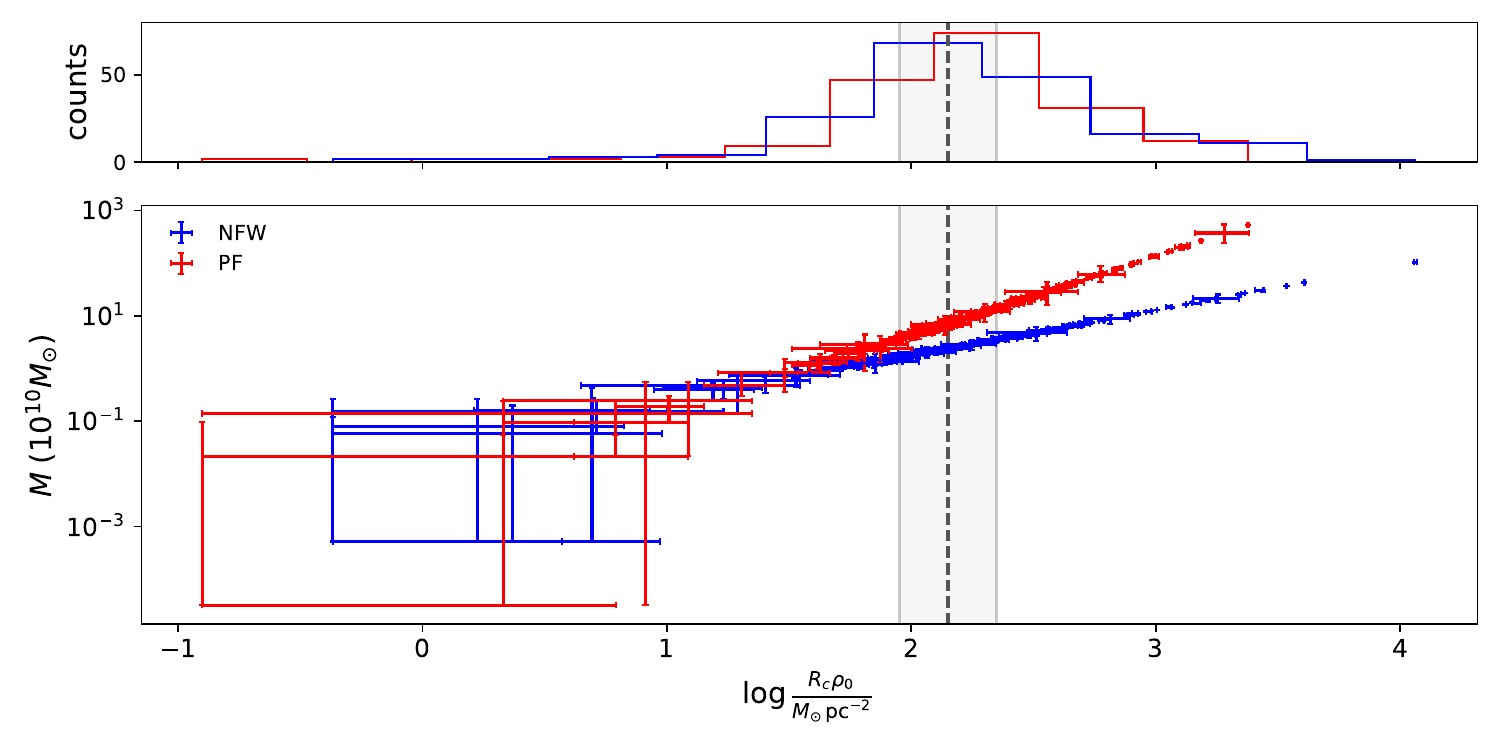}    
    \caption{Total mass as a function of the central surface density $\mu_{0}=R_{c}\rho_{0}$. The shaded region corresponds to the range of values reported in \cite{Donato:2009ab}. In red we present the perfect fluid halo results, and in blue the NFW case. In the upper panel we show the histogram of galaxy counts as a function of $\mu_{0}$.}
    \label{fig:salucci_relation}
\end{figure*}

The existence of universal relations between the central density and size of the DM halos of galaxies, their core radius or the total mass enclosed at some typical scales, among other relations, may provide relevant information in order to test dark matter candidates. For instance, one scaling relation of interest for dark matter halos is the nearly constant value found in \cite{Donato:2009ab, Gentile:2009bw}. It was found that

\begin{equation}
    \log{\left(\frac{R_c\rho_{0}}{M_{\cdot}pc^{-2}}\right)}= 2.15 \pm 0.2, \label{surface_constant}
\end{equation}

where $R_c$ is the core radius and the product $\mu_{0}=R_{c}\rho_{0}$ is called the central surface density. \\

Next, we explore if the resulting halos obtained with with our core-halo models, with fixed fermion mass and sef-interacting parameters as explained above, will have the same structural relations for $\mu_0$ as Eq. \eqref{surface_constant}. \\

In our case, we proceed as follows: we compute the core radius $R_c$ as defined by \cite{Donato:2009ab}, this is where the density of the system is one fourth of the central density, as they assumed that the DM halos follow a Burkert density profile. In the core-halo model we construct here, we adopt the same definition for $R_c$, although a more convenient definition for our core could be the transition point of the semi-degenerate system. Since we have fixed $m_f$ and $y$ with the best fit points, i.e. Eqs. \eqref{bfp_nfw} and \eqref{bfp_pf}, the only free parameter is $\rho_0$, thus we compute $\mu_0$ with the best fit point for each galaxy. The results are shown in figure \ref{fig:salucci_relation}. Note that of the 182 galaxies analyzed in this work, nearly $\sim 35\%$ fall between the ranges presented in \cite{Donato:2009ab, Gentile:2009bw}, for either of the core-halo cases. In the upper panel of Fig. \ref{fig:salucci_relation}, we have plotted a histogram that counts the number of galaxies where $\mu_0$ satisfies Eq. \ref{surface_constant}. It is interesting that the average coincides with the relation found in \cite{Donato:2009ab, Gentile:2009bw} but with a wider range for $\mu_0$. 
The lower panel of Figure \ref{fig:salucci_relation} shows the total mass $M$ of the core-halo configuration as a function of $\log(\mu_0)$. Unlike the constant values reported, in our model there is a clear exponential relation between $\mu_{0}$ and $M$. We find for the perfect fluid halo
\begin{equation}
M=2.28\times10^{-6}\mu_{0}^{3.54}\,,\label{Msurface:fluid}
\end{equation}
and for the NFW halo:
\begin{equation}
M=1.26\times10^{-4}\mu_{0}^{2.19}\,.\label{Msurface:nfw}
\end{equation} 

Future analysis or measurements of $\mu_0$ can then be compared with this functional dependence of the core-halo mass.\\

\begin{figure*}
\label{comparison models}
     \begin{subfigure}[b]{0.45\textwidth}
         \centering
         \includegraphics[width=\textwidth]{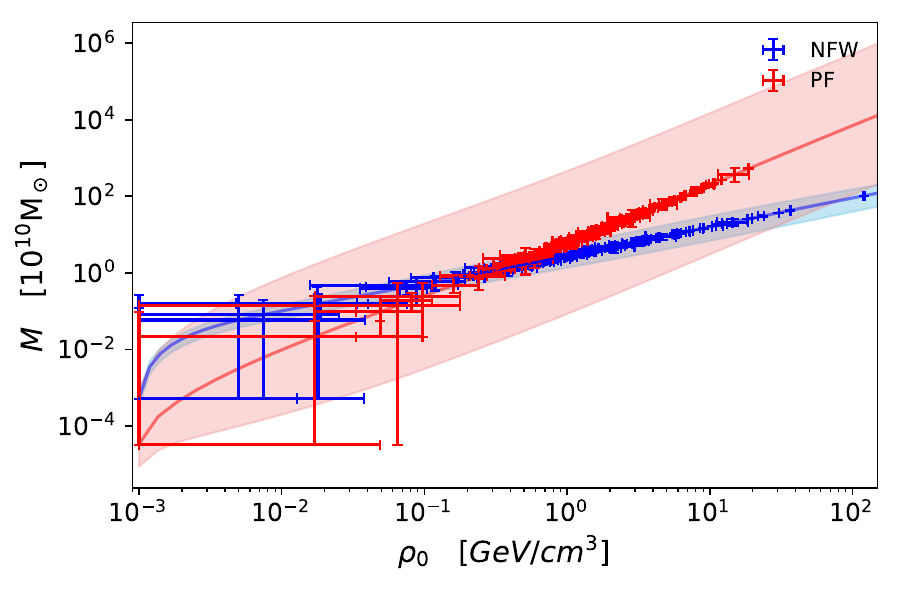}
         \caption{}
     \end{subfigure}
     \begin{subfigure}[b]{0.45\textwidth}
         \centering
         \includegraphics[width=\textwidth]{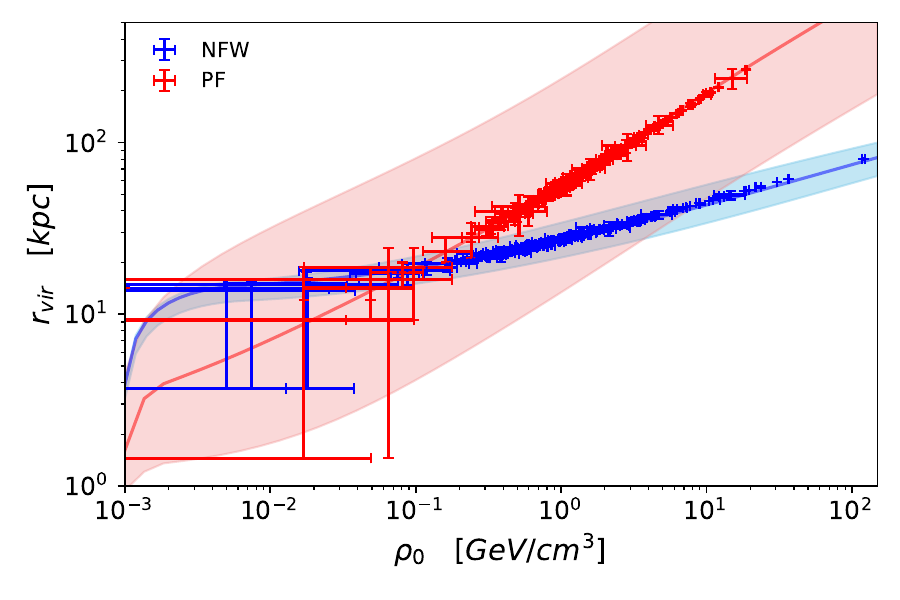}
         \caption{}
     \end{subfigure}
        \caption{Relations between the central density $\rho_{0}$  and the total mass (left) and virial radius (right) of the dark matter halos, for the perfect-fluid (red) and the NFW profile (blue) cases. Shaded regions represent the 68\% confidence level for the values fermion mass and self-interacting parameter obtained from fitting the rotation curves of the LSB galaxies, the solid line represents the minimum. Error bars indicate the range of $\rho_{0}$ values for each galaxy at the 68\% level and their corresponding total masses (left) or virial radii (right).}
        \label{fig:three graphs}
\end{figure*}

\begin{figure}
    \centering
    \includegraphics[width=\linewidth]{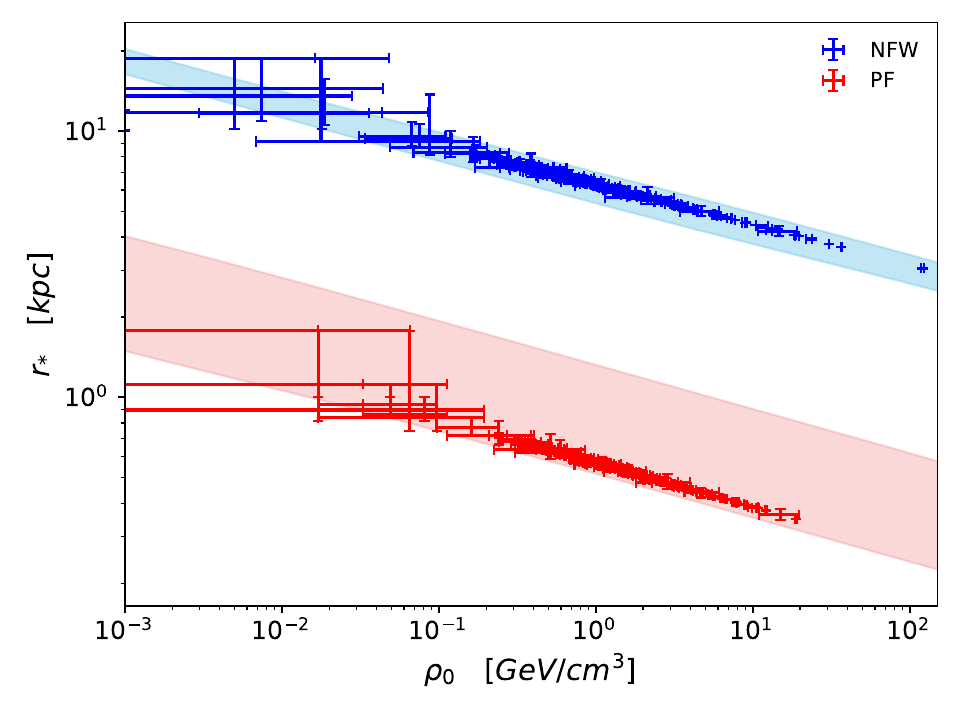}
    \caption{Relation between the central density $\rho_{0}$  and transition radius $r_{*}$ between the degenerate and halo regions of the dark matter halos, for the perfect-fluid (red) and the NFW profile (blue) cases. Shaded regions represent the 68\% confidence level for the values fermion mass and self-interacting parameter obtained from fitting the rotation curves of the LSB galaxies. Error bars indicate the range of $\rho_{0}$ values for each galaxy at the 68\% level and their corresponding degenerate core radii.}
    \label{fig:core_radius}
\end{figure}

Our analysis relies on the hypothesis that a self-interacting fermion can be described as a perfect fluid with EOS given by Eq. \eqref{polytrope} and two models for the halo. The free parameters of the EOS fixed by the LSB data. The values found for $m_f$ and $y$ are in a region that agrees with the full expressions for the pressure $p$ and mass density given by Eqs. \eqref{pressure} and \eqref{density} as can be seen in Fig. \ref{relative_difference}.\\

The structural properties of the core-halo are then a consequence of hydrostatic equilibrium. This analysis does not differ to that of polytropic self-gravitating objects: given a specific EOS, there is a unique family of self-gravitating objects. The characteristic size of the system is defined by the virial radius $r_{\text{vir}}$, which is the point where $\rho(r_{\text{vir}})=200\rho_c$, where $\rho_c$ is the critical density of the universe. There are interesting results that can be extracted from Figure \ref{fig:salucci_relation}:
\begin{enumerate}
\item The scaling relation found in \cite{Donato:2009ab,Gentile:2009bw} has been confirmed for both of our core-halo models. Most of the galaxies that we have analyzed have a central surface density $\mu_0$ that falls within the range parametrized by Eq. \eqref{surface_constant}. This result is relevant, as it is obtained after modifying the model for the dark matter pressure and applying it to two halo models as those considered in \cite{Donato:2009ab,Gentile:2009bw}. This suggests that the relation Eq. \ref{surface_constant} might be a true scaling relation that all models for dark matter should satisfy. 
\item Lower panel indicates that although the constant central density surface density might be true, this relation does not imply a universal halo mass. Furthermore, as shown by the Eqs. \ref{Msurface:fluid} and \ref{Msurface:nfw}, the total mass of the dark matter core-halo is a power law of the  central density surface density $\mu_0$. Combining rotation curve analysis with other observables, the gravitational lensing of the halo \cite{Nunez:2010ug} for instance, could be used to distinguish between the perfect fluid or the NFW halo models, because they predict different halo masses.
\end{enumerate}

In our case, the remaining 179 galaxies that we have fitted by varying only $\rho_0$ belong to a family of core-halos. This fact is illustrated in Figure \ref{fig:three graphs}. Left panel shows the typical plot $M$  versus $\rho_0$ for the two halo models while the right panel shows the virial radius of the core-halo system $r_{\text{vir}}$ vs. $\rho_0$.\\

All galaxies can be seen as a particular realization of a self-gravitating object which is a solution of the TOV system with a EOS given by Eq. \eqref{polytrope}. 
Shaded regions represent the 68\% confidence level for the values fermion mass and self-interacting parameter obtained from fitting the rotation curves of the LSB galaxies, the solid line represents the minimum. Error bars indicate the range of $\rho_{0}$ values for each galaxy at the 68\%.
There are important differences that depend on the halo model under consideration. The NFW halos are smaller and less massive than the perfect fluid ones, this is expected due to the fact that the density profile behaves as $\rho\sim r^{-2}$ for an isothermal gas sphere and as $\rho\sim r^{-3}$ for the NFW density profile.\\
\begin{figure}
    \centering
    \includegraphics[width=\linewidth]{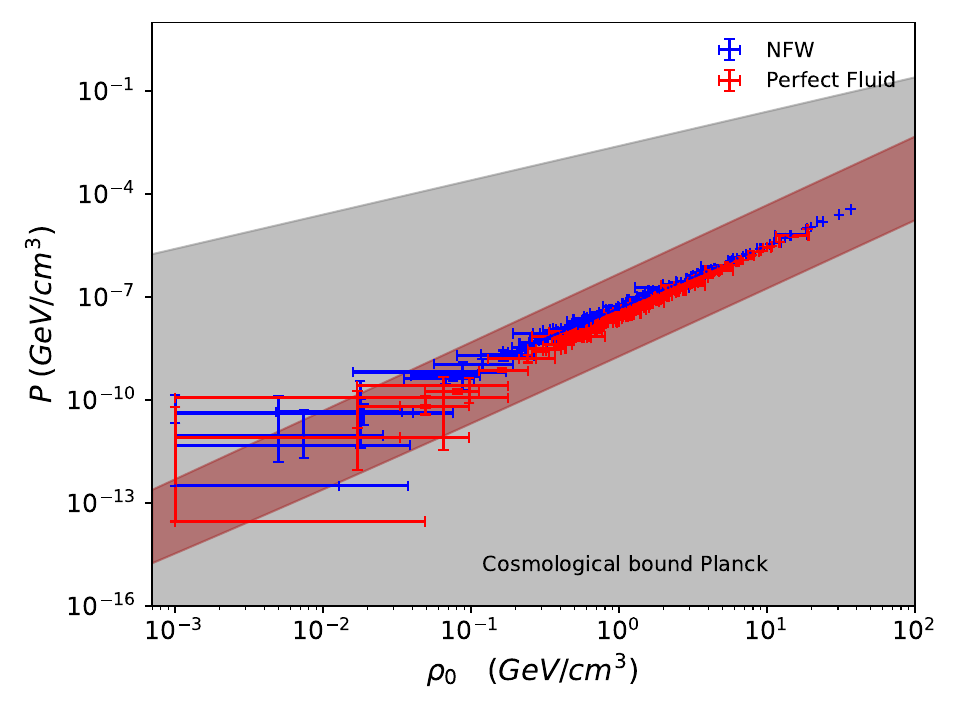}
    \caption{Shaded gray region represents the current cosmological bound for a barotropic EOS for dark matter \cite{Xu:2013mqe} and it is compared for the pressure obtained for the EOS given by Eq. \eqref{polytrope} with $m_f$ and $y$ given by Eqs. \eqref{bfp_nfw} for the NFW halo and \eqref{bfp_pf} for the perfect fluid halo (brown shaded region). Data points correspond to the pressure at the center of each galaxy taken from Table \ref{table_sparc}.     }
    \label{fig:eosfinal}
\end{figure}\\
A cosmological model with an equation of state as given by Eq.  \eqref{polytrope} lies beyond the analysis presented here.  To our knowledge, the research beyond the pressure-less hypothesis for dark matter is just starting. Current cosmological bounds for a barotropic equation of state are consistent with a pressure less dark matter EOS, although significant room for $p \ne 0$ is allowed \cite{Muller:2004yb,Calabrese:2009zza,Kumar:2012gr,Armendariz-Picon:2013jej, Xu:2013mqe}.   
At galactic scales, dark matter pressure could be required to understand some of the observation and alleviate tensions with the current CDM paradigm \cite{Serra:2011jh,Barranco:2013wy,Acena:2021wjx, Boshkayev:2021wns}. This work naturally extends the analysis for a fermionic dark matter particle that in the degenerate limit satisfies a polytropic effective EOS \cite{Randall2017,Destri2013,Destri2013a,Domcke2015,barranco2019constraining,deVega:2013ysa}. 
We found that if the self-interaction parameter $y$ is zero, it is unlike that a single fermion can explain the diversity of halos. On the other hand, a single fermion with a significant contribution of self-interaction can explain a diversity of galaxies. \\

Note that the effective pressure obtained for the center of all the galaxies analyzed in this work fall below the cosmological limit for a barotropic equation fo state for a halo modeled as a perfect fluid and just in the limit of testability for the NFW halos. This is illustrated in Figure \ref{fig:eosfinal} where we have overlapped the limits obtained in \cite{Xu:2013mqe} against the pressure obtained for the EOS \eqref{polytrope} with the values for $m_f$ and $y$ as expressed in Eqs. \eqref{bfp_nfw} and \eqref{bfp_pf}. Furthermore, the data points correspond to the value of the pressure for the central density that explains the data for each of the 185 galaxies analyzed here. 
In can be inferred from Fig. \ref{fig:eosfinal} that galactic analysis could be more sensible to the pressure-less hypothesis for dark matter.

To calculate the self-interacting cross-section in a mean field theory with a Lagrangian given by $\mathcal{L}_{\mathrm{int}}=g\bar{\chi}\gamma_{\mu}\chi\mathbf{V}^{\mu} $, as shown in \cite{Girmohanta2022}, the cross-section is written as 

\begin{equation}
    \sigma=\frac{1}{4\pi}\frac{y^4}{m_f^2}\left[ \frac{1}{1 + r} - \frac{\ln{(1+r)}}{r(2+r)}\right],
\end{equation}

where r is the defined as

\begin{equation}
    r=\left(\frac{\beta_{rel}m_{f}}{m_{I}}\right)^2
\end{equation}

and $\beta_{rel}$ is the relative velocity between particles in the non-relativistic limit. In our model the self-interacting parameter $y$ includes in its definition both the coupling constant $g$ and the mass of the particle that mediates the interaction, $y=gm_f/m_I$, as a result the cross-section cannot be uniquely determined from $m_f$ and $y$ alone. Nevertheless, rewriting the expression in terms of the coupling constant $g$

\begin{equation}
    \sigma=\frac{1}{4\pi}\frac{y^4}{m_f^2}g^2\left[ \frac{1}{g^2 + y^2\beta_{rel}^2} - \frac{g^2\left(\ln{(g^2+y^2\beta_{rel}^2)-\ln{g^2}}\right)}{y^2\beta_{rel}^2(2g^2+y^2\beta_{rel}^2)}\right],
\end{equation}

we can estimate what values of the coupling constant and the mediator mass would be consistent with the constraints of the bullet cluster \cite{Markevitch2004ApJ}. With  $\sigma/m < 1 \mathrm{cm}^2 /\mathrm{g}$, and considering that in a cluster $v_{rel} \sim 3\times10^3 km/s$ \cite{TULIN20181}, which corresponds to $\beta_{rel} \sim 10^{-2}$, for DM particles in galaxy clusters. Using our results, to be consistent with the bullet cluster constraints, the coupling constant should be $g\sim \mathcal{O}(10^{-13})$ in the case of the NFW profile and $g\sim \mathcal{O}(10^{-14})$ in the perfect fluid case, which would entail a mediator with a mass of $m_H\sim \mathcal{O}(10^{-13}\mathrm{eV})$ and $m_H\sim \mathcal{O}(10^{-15}\mathrm{eV})$ respectively.

It is important to point out that the self-interacting regime is confined to the degenerate core of our model, with radii below 2kpc or 20kpc depending on the halo type as seen in Fig. \ref{fig:core_radius}. Outside this region the model is described by a perfect fluid or NFW profile, which, as it dilutes,  becomes an effectively pressure-less EoS at the outskirts of the halo, equivalent to CDM.

As a consequence, a direct comparison to constraints obtained through simulations at cluster scales is not straightforward. Because of the size of the degenerate cores, it is not clear if the simulations would be sensitive to the degenerate region of the halo.

\section{Conclusions}\label{conclusions}

Within certain conditions the quantum nature of the particle that conforms the dark matter must be taken into account, and the pressure-less CDM paradigm must be challenged. For the case of a gas of fermionic particles this means becoming degenerate below a certain temperature. In the simplest model, this degenerate core depends on three parameters: the fermion mass, the central density of the system and if the particle has self-interactions and its strength. Of these parameters, the two relevant ones to describe the particle that would make up dark matter are the fermion mass and the self-interaction parameter, since these must be universal and fundamental, while the central density is a parameter that changes from system to system. \\

Outside of this region of degeneracy the system must have a halo, we chose a perfect fluid like halo, and a NFW type halo. Using six LSB galaxies we show that the inclusion of self-interactions yields better results when fitting the rotational curves. If the fermion has a perfect fluid halo, the mass must fall within the range of $155.10~\rm{eV}< m_{f} < 313.26~\rm{eV}$ and $1465.31 < y <6002.04$ for the self-interaction parameter. If the halo is of the NFW type, then the values are $42.31 ~\rm{eV} < m_{f} <49.23 ~\rm{eV}$ and $71.43< y <  132.65$.\\

Fixing the fermion mass and self interacting parameter to the values that minimize the combined $\chi^{2}$ of the LSB galaxies, there remains only one free parameter, the central density, a value that depends on the system studied. We tested the $m_{f}$ and $y$ parameters in two different cases: the Milky Way rotational curve and the SPARC dataset. The Milky Way's rotation curve can be decomposed into four components apart from the dark matter halo —the inner bulge, the main bulge, the disk, and the black hole. Of these baryonic components, all but the black hole are described by an exponential sphere model. Performing the central density fitting for the Milky Way, the best fit for the NFW halo case is $\chi^{2}/dof=0.94$, while it is $\chi^{2}/dof=0.79$ for the perfect fluid. The second case to test the core-halo model with a self-interacting degenerate fermion core was the SPARC dataset. This dataset includes 175 spiral and irregular galaxies and it includes the baryonic contributions of the gas, disk, and bulge of each galaxy. If the outer halo is of the perfect fluid type, the median $\chi^{2}/dof$ is 2.85; for the NFW case, it is 6.09. Overall the case with a perfect fluid halo provided better fits for the LSB galaxies, the Milky Way and SPARC dataset.\\
By fixing $m_f$ and $y$, there is only one free parameter, the central density $\rho_0$. This values are reported in Tables |\ref{SI best fit},\ref{perfect fluid fixed}, \ref{NFW fixed} and \ref{table_sparc}. The matching condition of the degenerate fermion core and the halo determines $\rho_s$ and $r_s$ once $\rho_0$ is fitted by data. Since $\rho_s < \rho_0$, we can see that $\rho_s$ is in reasonable agreement with previous fit that used NFW profiles only for the dark matter halo \cite{Jimenez:2002vy}.

The halos found in the analysis have certain structural properties that are interesting, such as the total mass of the halo having an exponential dependency to the central surface density, instead of the nearly constant value reported.

Our core-halo model predicts a core made of degenerate fermions. This condition of degeneracy is system dependent.  
Outside the degenerate core it becomes dilute and with negligible pressure at cosmological scales, behaving as a pressure-less fluid at large scales. Furthermore, as it can be noted from Figure \ref{fig:eosfinal}, the central pressure where the fermions are degenerated is much smaller that the current limit for the pressure for dark matter obtained from PLANCK data \cite{Xu:2013mqe} by assuming a barotropic equation of state for dark matter. 
Because of this we expect the model to reproduce the standard CDM results at large scales. 

A proper cosmological simulation would require an analysis to calculate if these conditions are met at any moment in the evolution of the universe. Which clearly is far beyond the scope of this work.

\appendix
\section{Results from fits to SPARC dataset}

\begin{table*}
\caption{Summary of the fits for the SPARC catalog. The first column refers to the name of the galaxy.  The two cases for the halo model are reported: the NFW halo and the perfect fluid halo. For each case we report: the first column is the best fit value for the central density $\rho_0$ in units of GeV/cm$^3$ obtained form the fit as explained in Section \ref{sec:sparc}. The second column shows the minimum value of the $\chi¨2$ function and the third column the f¡goodness of the fit expressed with the value of the minimum of the $\chi^2$ function divided by the degree of freedom. For completeness, some properties of each galaxy under analysis are reported as: $Q$ the quality flag: 1 = High, 2= Medium, 3= Low. The Hubble type obtained from \cite{deVancoleurs1991}, \cite{Schombert1992}  or the NASA/IPAC Extragalactic Database (NED) and finally the number of data points $N_{data}$. }\label{table_sparc}
\begin{ruledtabular}
\begin{tabular}{l|ccc|ccc|ccc}
& \multicolumn{3}{c|}{NFW Halo} & \multicolumn{3}{c|}{Perfect Fluid Halo} \\
Galaxy & \begin{tabular}{c}
$\rho_0$ \\
$[\mathrm{GeV/cm^3}]$
\end{tabular} & $\chi^2$ & $\chi^2/\mathrm{dof}$ & \begin{tabular}{c}
$\rho_0$ \\
$[\mathrm{GeV/cm^3}]$
\end{tabular} & $\chi^2$ & $\chi^2/\mathrm{dof}$ & Q & Hubble Type & $N_{data}$\\
\hline
CamB & 0.075 & 32.268 & 4.034 & 0.081 & 33.266 & 4.158 & 2 & Im & 9 \\
D512-2 & 0.386 & 11.618 & 3.873 & 0.497 & 6.625 & 2.208 & 2 & Im & 4 \\
D564-8 & 0.192 & 9.990 & 1.998 & 0.241 & 4.942 & 0.988 & 2 & Im & 6 \\
D631-7 & 0.300 & 10.246 & 0.683 & 0.481 & 34.920 & 2.328 & 1 & Im & 16 \\
DDO064 & 0.915 & 11.258 & 0.866 & 1.042 & 7.938 & 0.611 & 1 & Im & 14 \\
DDO154 & 0.332 & 1272.398 & 115.673 & 0.497 & 296.096 & 26.918 & 2 & Im & 12 \\
DDO161 & 0.248 & 276.003 & 9.200 & 0.497 & 219.345 & 7.312 & 1 & Im & 31 \\
DDO168 & 0.610 & 55.513 & 6.168 & 0.786 & 34.772 & 3.864 & 2 & Im & 10 \\
DDO170 & 0.233 & 50.609 & 7.230 & 0.465 & 13.737 & 1.962 & 2 & Im & 8 \\
ESO079-G014 & 3.249 & 601.878 & 42.991 & 3.187 & 194.939 & 13.924 & 1 & Sbc & 15 \\
ESO116-G012 & 1.135 & 34.247 & 2.446 & 1.714 & 12.915 & 0.923 & 1 & Sd & 15 \\
ESO444-G084 & 1.052 & 131.856 & 21.976 & 1.346 & 77.474 & 12.912 & 2 & Im & 7 \\
ESO563-G021 & 6.280 & 2989.194 & 103.076 & 7.815 & 1281.477 & 44.189 & 1 & Sbc & 30 \\
F561-1 & 0.067 & 5.805 & 1.161 & 0.161 & 2.617 & 0.523 & 3 & Sm & 6 \\
F563-1 & 1.500 & 18.871 & 1.179 & 1.730 & 7.813 & 0.488 & 1 & Sm & 17 \\
F563-V1 & 0.018 & 2.960 & 0.592 & 0.049 & 1.441 & 0.288 & 3 & Im & 6 \\
F563-V2 & 1.814 & 8.381 & 0.931 & 2.403 & 5.454 & 0.606 & 1 & Im & 10 \\
F565-V2 & 0.530 & 0.519 & 0.086 & 0.914 & 1.394 & 0.232 & 2 & Im & 7 \\
F567-2 & 0.172 & 3.000 & 0.750 & 0.337 & 0.969 & 0.242 & 3 & Sm & 5 \\
F568-1 & 1.932 & 1.895 & 0.172 & 2.467 & 0.620 & 0.056 & 1 & Sc & 12 \\
F568-3 & 0.737 & 28.939 & 1.702 & 1.026 & 44.656 & 2.627 & 1 & Sd & 18 \\
F568-V1 & 1.598 & 8.052 & 0.575 & 2.227 & 3.646 & 0.260 & 1 & Sd & 15 \\
F571-8 & 1.481 & 321.556 & 26.796 & 2.147 & 257.761 & 21.480 & 1 & Sc & 13 \\
F571-V1 & 0.530 & 4.225 & 0.704 & 0.882 & 1.902 & 0.317 & 2 & Sd & 7 \\
F574-1 & 0.892 & 13.675 & 1.052 & 1.330 & 1.793 & 0.138 & 1 & Sd & 14 \\
F574-2 & 0.005 & 0.294 & 0.074 & 0.017 & 0.354 & 0.088 & 3 & Sm & 5 \\
F579-V1 & 1.538 & 38.637 & 2.972 & 1.938 & 31.811 & 2.447 & 1 & Sc & 14 \\
F583-1 & 0.580 & 9.013 & 0.376 & 0.898 & 6.757 & 0.282 & 1 & Sm & 25 \\
F583-4 & 0.354 & 15.247 & 1.386 & 0.641 & 8.713 & 0.792 & 1 & Sc & 12 \\
IC2574 & 0.163 & 172.003 & 5.212 & 0.321 & 494.733 & 14.992 & 2 & Sm & 34 \\
IC4202 & 3.332 & 1793.212 & 57.846 & 4.196 & 1272.909 & 41.062 & 1 & Sbc & 32 \\
KK98-251 & 0.278 & 10.341 & 0.739 & 0.337 & 4.582 & 0.327 & 2 & Im & 15 \\
NGC0024 & 1.481 & 342.594 & 12.235 & 1.874 & 273.310 & 9.761 & 1 & Sc & 29 \\
NGC0055 & 0.372 & 42.817 & 2.141 & 0.657 & 78.886 & 3.944 & 2 & Sm & 21 \\
NGC0100 & 0.625 & 6.080 & 0.304 & 0.994 & 4.257 & 0.213 & 1 & Scd & 21 \\
NGC0247 & 0.756 & 468.670 & 18.747 & 1.122 & 198.396 & 7.936 & 2 & Sd & 26 \\
NGC0289 & 6.863 & 760.733 & 28.175 & 3.876 & 136.736 & 5.064 & 2 & Sbc & 28 \\
NGC0300 & 0.859 & 55.081 & 2.295 & 1.266 & 17.196 & 0.717 & 2 & Sd & 25 \\
NGC0801 & 9.070 & 617.703 & 51.475 & 4.612 & 137.948 & 11.496 & 1 & Sc & 13 \\
NGC0891 & 2.250 & 830.057 & 48.827 & 2.691 & 530.421 & 31.201 & 1 & Sb & 18 \\
\end{tabular}
\end{ruledtabular}
\end{table*}

\begin{table*}
\caption{SPARC fits comparison (part 2 of 5)}
\label{SPARC_table_2}
\begin{ruledtabular}
\begin{tabular}{l|ccc|ccc|ccc}
& \multicolumn{3}{c|}{NFW Halo} & \multicolumn{3}{c|}{Perfect Fluid Halo} \\
Galaxy & \begin{tabular}{c}
$\rho_0$ \\
$[\mathrm{GeV/cm^3}]$
\end{tabular} & $\chi^2$ & $\chi^2/\mathrm{dof}$ & \begin{tabular}{c}
$\rho_0$ \\
$[\mathrm{GeV/cm^3}]$
\end{tabular} & $\chi^2$ & $\chi^2/\mathrm{dof}$ & Q & Hubble Type & $N_{data}$\\
\hline
NGC1003 & 1.288 & 2708.988 & 77.400 & 1.490 & 796.191 & 22.748 & 1 & Scd & 36 \\
NGC1090 & 4.349 & 849.573 & 36.938 & 2.931 & 137.763 & 5.990 & 1 & Sbc & 24 \\
NGC1705 & 1.224 & 212.471 & 16.344 & 1.682 & 146.950 & 11.304 & 3 & BCD & 14 \\
NGC2366 & 0.340 & 157.553 & 6.302 & 0.513 & 55.728 & 2.229 & 3 & Im & 26 \\
NGC2403 & 1.500 & 8359.532 & 116.105 & 2.067 & 2061.135 & 28.627 & 1 & Scd & 73 \\
NGC2683 & 3.418 & 55.956 & 5.596 & 3.348 & 6.141 & 0.614 & 2 & Sb & 11 \\
NGC2841 & 30.421 & 7144.075 & 145.797 & 12.074 & 299.863 & 6.120 & 1 & Sb & 50 \\
NGC2903 & 3.128 & 2398.145 & 72.671 & 3.315 & 756.719 & 22.931 & 1 & Sbc & 34 \\
NGC2915 & 0.766 & 67.022 & 2.311 & 1.218 & 42.857 & 1.478 & 2 & BCD & 30 \\
NGC2955 & 1.372 & 258.424 & 11.236 & 2.259 & 190.191 & 8.269 & 1 & Sb & 24 \\
NGC2976 & 2.367 & 11.541 & 0.444 & 2.515 & 9.388 & 0.361 & 2 & Sc & 27 \\
NGC2998 & 13.266 & 514.322 & 42.860 & 5.333 & 93.259 & 7.772 & 1 & Sc & 13 \\
NGC3109 & 0.473 & 52.548 & 2.190 & 0.738 & 6.499 & 0.271 & 1 & Sm & 25 \\
NGC3198 & 1.957 & 7290.156 & 173.575 & 2.547 & 1214.434 & 28.915 & 1 & Sc & 43 \\
NGC3521 & 3.049 & 123.506 & 3.088 & 3.700 & 47.638 & 1.191 & 1 & Sbc & 41 \\
NGC3726 & 1.224 & 163.382 & 14.853 & 1.698 & 84.163 & 7.651 & 2 & Sc & 12 \\
NGC3741 & 0.300 & 153.944 & 7.697 & 0.497 & 54.644 & 2.732 & 1 & Im & 21 \\
NGC3769 & 1.179 & 62.454 & 5.678 & 1.586 & 18.779 & 1.707 & 2 & Sb & 12 \\
NGC3877 & 1.272 & 51.097 & 4.258 & 1.794 & 53.192 & 4.433 & 2 & Sc & 13 \\
NGC3893 & 2.721 & 30.301 & 3.367 & 3.155 & 6.360 & 0.707 & 1 & Sc & 10 \\
NGC3917 & 1.462 & 71.463 & 4.466 & 1.986 & 24.507 & 1.532 & 1 & Scd & 17 \\
NGC3949 & 1.079 & 2.916 & 0.486 & 1.586 & 3.527 & 0.588 & 2 & Sbc & 7 \\
NGC3953 & 3.088 & 11.297 & 1.614 & 3.460 & 1.275 & 0.182 & 1 & Sbc & 8 \\
NGC3972 & 1.304 & 17.584 & 1.954 & 1.954 & 7.421 & 0.825 & 1 & Sbc & 10 \\
NGC3992 & 18.211 & 254.943 & 31.868 & 7.815 & 17.645 & 2.206 & 1 & Sbc & 9 \\
NGC4010 & 0.975 & 27.491 & 2.499 & 1.506 & 30.611 & 2.783 & 2 & Sd & 12 \\
NGC4013 & 3.596 & 1184.242 & 33.835 & 3.091 & 341.036 & 9.744 & 2 & Sb & 36 \\
NGC4051 & 0.837 & 2.376 & 0.396 & 1.266 & 3.410 & 0.568 & 2 & Sbc & 7 \\
NGC4068 & 0.456 & 0.964 & 0.193 & 0.513 & 1.137 & 0.227 & 2 & Im & 6 \\
NGC4085 & 0.737 & 23.954 & 3.992 & 1.138 & 27.425 & 4.571 & 2 & Sc & 7 \\
NGC4088 & 0.610 & 84.855 & 7.714 & 1.074 & 62.769 & 5.706 & 1 & Sbc & 12 \\
NGC4100 & 3.550 & 135.071 & 5.873 & 3.460 & 14.742 & 0.641 & 1 & Sbc & 24 \\
NGC4138 & 2.427 & 1.805 & 0.301 & 2.787 & 5.116 & 0.853 & 2 & S0 & 7 \\
NGC4157 & 2.619 & 174.583 & 10.911 & 2.611 & 70.499 & 4.406 & 1 & Sb & 17 \\
NGC4183 & 1.209 & 57.222 & 2.601 & 1.490 & 10.818 & 0.492 & 1 & Scd & 23 \\
NGC4214 & 1.256 & 90.956 & 6.997 & 1.746 & 65.685 & 5.053 & 2 & Im & 14 \\
NGC4217 & 0.450 & 730.019 & 40.557 & 0.978 & 720.999 & 40.055 & 1 & Sb & 19 \\
NGC4389 & 0.001 & 83.915 & 16.783 & 0.001 & 83.899 & 16.780 & 3 & Sbc & 6 \\
NGC4559 & 1.135 & 194.364 & 6.270 & 1.458 & 58.549 & 1.889 & 1 & Scd & 32 \\
NGC5005 & 2.138 & 14.179 & 0.834 & 2.851 & 13.325 & 0.784 & 1 & Sbc & 18 \\
\end{tabular}
\end{ruledtabular}
\end{table*}

\begin{table*}
\caption{SPARC fits comparison (part 3 of 5)}
\label{SPARC_table_3}
\begin{ruledtabular}
\begin{tabular}{l|ccc|ccc|ccc}
& \multicolumn{3}{c|}{NFW Halo} & \multicolumn{3}{c|}{Perfect Fluid Halo} \\
Galaxy & \begin{tabular}{c}
$\rho_0$ \\
$[\mathrm{GeV/cm^3}]$
\end{tabular} & $\chi^2$ & $\chi^2/\mathrm{dof}$ & \begin{tabular}{c}
$\rho_0$ \\
$[\mathrm{GeV/cm^3}]$
\end{tabular} & $\chi^2$ & $\chi^2/\mathrm{dof}$ & Q & Hubble Type & $N_{data}$\\
\hline
NGC5033 & 10.427 & 1884.089 & 89.719 & 5.381 & 260.485 & 12.404 & 1 & Sc & 22 \\
NGC5055 & 1.558 & 34157.836 & 1265.105 & 2.435 & 14853.161 & 550.117 & 1 & Sbc & 28 \\
NGC5371 & 3.642 & 521.579 & 28.977 & 2.979 & 171.696 & 9.539 & 1 & Sbc & 19 \\
NGC5585 & 0.701 & 125.189 & 5.443 & 1.074 & 126.406 & 5.496 & 1 & Sd & 24 \\
NGC5907 & 13.266 & 1999.066 & 111.059 & 5.669 & 210.600 & 11.700 & 1 & Sc & 19 \\
NGC5985 & 19.650 & 2420.273 & 75.634 & 10.681 & 86.643 & 2.708 & 1 & Sb & 33 \\
NGC6015 & 2.619 & 2017.190 & 46.911 & 3.107 & 855.543 & 19.896 & 2 & Scd & 44 \\
NGC6195 & 1.372 & 457.251 & 20.784 & 2.243 & 273.580 & 12.435 & 1 & Sb & 23 \\
NGC6503 & 1.519 & 1356.585 & 45.219 & 1.762 & 121.255 & 4.042 & 1 & Scd & 31 \\
NGC6674 & 120.202 & 2314.713 & 144.670 & 9.032 & 193.781 & 13.842 & 1 & Sb & 15 \\
NGC6674 & 36.553 & 2052.109 & 146.579 & 9.032 & 193.781 & 13.842 & 1 & Sb & 15 \\
NGC6789 & 14.681 & 0.197 & 0.066 & 15.004 & 0.350 & 0.117 & 2 & BCD & 4 \\
NGC6946 & 1.408 & 535.461 & 9.394 & 1.970 & 276.056 & 4.843 & 1 & Scd & 58 \\
NGC7331 & 5.820 & 3218.030 & 91.944 & 4.228 & 1321.483 & 37.757 & 1 & Sb & 36 \\
NGC7793 & 1.079 & 133.979 & 2.977 & 1.554 & 94.918 & 2.109 & 1 & Sd & 46 \\
NGC7814 & 5.675 & 433.148 & 25.479 & 4.869 & 65.531 & 3.855 & 1 & Sab & 18 \\
PGC51017 & 0.007 & 8.776 & 1.755 & 0.017 & 8.828 & 1.766 & 3 & BCD & 6 \\
UGC00128 & 3.462 & 8898.012 & 423.715 & 2.387 & 1308.150 & 62.293 & 1 & Sdm & 22 \\
UGC00191 & 0.504 & 368.155 & 46.019 & 0.914 & 215.314 & 26.914 & 1 & Sm & 9 \\
UGC00634 & 1.256 & 181.658 & 60.553 & 1.506 & 45.429 & 15.143 & 2 & Sm & 4 \\
UGC00731 & 0.450 & 129.682 & 11.789 & 0.786 & 54.320 & 4.938 & 1 & Im & 12 \\
UGC00891 & 0.308 & 8.025 & 2.006 & 0.577 & 16.209 & 4.052 & 2 & Sm & 5 \\
UGC01230 & 1.372 & 13.972 & 1.397 & 1.618 & 1.878 & 0.188 & 1 & Sm & 11 \\
UGC01281 & 0.492 & 22.302 & 0.929 & 0.690 & 6.394 & 0.266 & 1 & Sdm & 25 \\
UGC02023 & 0.386 & 0.378 & 0.094 & 0.513 & 0.890 & 0.223 & 2 & Im & 5 \\
UGC02259 & 0.870 & 321.838 & 45.977 & 1.362 & 173.835 & 24.834 & 2 & Sdm & 8 \\
UGC02455 & 0.001 & 188.327 & 26.904 & 0.001 & 188.311 & 26.902 & 3 & Im & 8 \\
UGC02487 & 25.000 & 18876.248 & 1179.766 & 18.751 & 132.382 & 8.274 & 1 & S0 & 17 \\
UGC02885 & 23.607 & 952.406 & 52.911 & 10.393 & 182.705 & 10.150 & 1 & Sc & 19 \\
UGC02916 & 1.000 & 2381.415 & 56.700 & 1.650 & 2295.946 & 54.665 & 2 & Sab & 43 \\
UGC02953 & 14.496 & 48482.022 & 425.281 & 8.792 & 4442.007 & 38.965 & 2 & Sab & 115 \\
UGC03205 & 8.956 & 2446.804 & 52.060 & 5.381 & 287.954 & 6.127 & 1 & Sab & 48 \\
UGC03546 & 4.240 & 941.351 & 32.460 & 3.428 & 273.325 & 9.425 & 1 & Sa & 30 \\
UGC03580 & 1.092 & 4181.832 & 90.909 & 1.458 & 2630.997 & 57.196 & 2 & Sa & 47 \\
UGC04278 & 0.650 & 18.116 & 0.755 & 0.994 & 30.023 & 1.251 & 1 & Sd & 25 \\
UGC04305 & 0.019 & 57.956 & 2.760 & 0.049 & 53.090 & 2.528 & 3 & Im & 22 \\
UGC04325 & 1.304 & 265.886 & 37.984 & 1.906 & 173.012 & 24.716 & 1 & Sm & 8 \\
UGC04483 & 0.848 & 30.943 & 4.420 & 0.914 & 25.222 & 3.603 & 2 & Im & 8 \\
UGC04499 & 0.417 & 31.700 & 3.963 & 0.738 & 6.770 & 0.846 & 1 & Sdm & 9 \\
UGC05005 & 0.467 & 44.807 & 4.481 & 0.850 & 23.582 & 2.358 & 1 & Im & 11 \\
\end{tabular}
\end{ruledtabular}
\end{table*}

\begin{table*}
\caption{SPARC fits comparison (part 4 of 5)}
\label{SPARC_table_4}
\begin{ruledtabular}
\begin{tabular}{l|ccc|ccc|ccc}
& \multicolumn{3}{c|}{NFW Halo} & \multicolumn{3}{c|}{Perfect Fluid Halo} \\
Galaxy & \begin{tabular}{c}
$\rho_0$ \\
$[\mathrm{GeV/cm^3}]$
\end{tabular} & $\chi^2$ & $\chi^2/\mathrm{dof}$ & \begin{tabular}{c}
$\rho_0$ \\
$[\mathrm{GeV/cm^3}]$
\end{tabular} & $\chi^2$ & $\chi^2/\mathrm{dof}$ & Q & Hubble Type & $N_{data}$\\
\hline
UGC05253 & 4.460 & 6990.842 & 97.095 & 4.677 & 1953.278 & 27.129 & 2 & Sab & 73 \\
UGC05414 & 0.517 & 9.155 & 1.831 & 0.706 & 2.381 & 0.476 & 1 & Im & 6 \\
UGC05716 & 0.456 & 358.821 & 32.620 & 0.754 & 89.173 & 8.107 & 2 & Sm & 12 \\
UGC05721 & 1.012 & 426.984 & 19.408 & 1.554 & 303.699 & 13.804 & 1 & Sd & 23 \\
UGC05750 & 0.261 & 4.931 & 0.493 & 0.481 & 7.401 & 0.740 & 1 & Sdm & 11 \\
UGC05764 & 1.321 & 1647.441 & 183.049 & 1.554 & 1083.301 & 120.367 & 2 & Im & 10 \\
UGC05829 & 0.328 & 10.610 & 1.061 & 0.545 & 2.957 & 0.296 & 2 & Im & 11 \\
UGC05918 & 0.345 & 26.668 & 3.810 & 0.497 & 15.316 & 2.188 & 2 & Im & 8 \\
UGC05986 & 1.408 & 56.508 & 4.036 & 2.067 & 21.344 & 1.525 & 2 & Sm & 15 \\
UGC05999 & 0.658 & 28.378 & 7.095 & 1.026 & 19.645 & 4.911 & 2 & Im & 5 \\
UGC06399 & 0.692 & 7.966 & 0.996 & 1.122 & 0.873 & 0.109 & 1 & Sm & 9 \\
UGC06446 & 0.719 & 102.009 & 6.376 & 1.154 & 55.359 & 3.460 & 1 & Sd & 17 \\
UGC06614 & 1.982 & 217.732 & 18.144 & 2.403 & 127.994 & 10.666 & 1 & Sa & 13 \\
UGC06628 & 0.018 & 4.074 & 0.679 & 0.065 & 3.233 & 0.539 & 2 & Sm & 7 \\
UGC06667 & 0.796 & 43.025 & 5.378 & 1.250 & 9.597 & 1.200 & 1 & Scd & 9 \\
UGC06786 & 7.692 & 8929.987 & 202.954 & 6.374 & 901.169 & 20.481 & 1 & S0 & 45 \\
UGC06787 & 7.220 & 17358.013 & 247.972 & 6.230 & 4303.018 & 61.472 & 2 & Sab & 71 \\
UGC06818 & 0.367 & 9.193 & 1.313 & 0.609 & 19.099 & 2.728 & 2 & Sm & 8 \\
UGC06917 & 0.950 & 12.985 & 1.299 & 1.458 & 2.156 & 0.216 & 1 & Sm & 11 \\
UGC06923 & 0.692 & 5.077 & 1.015 & 1.026 & 2.063 & 0.413 & 2 & Im & 6 \\
UGC06930 & 0.950 & 12.594 & 1.399 & 1.362 & 1.604 & 0.178 & 1 & Sd & 10 \\
UGC06973 & 0.382 & 619.401 & 77.425 & 0.593 & 638.980 & 79.872 & 3 & Sab & 9 \\
UGC06983 & 1.106 & 29.848 & 1.865 & 1.618 & 10.021 & 0.626 & 1 & Scd & 17 \\
UGC07089 & 0.315 & 3.026 & 0.275 & 0.561 & 5.810 & 0.528 & 2 & Sdm & 12 \\
UGC07125 & 0.190 & 11.600 & 0.967 & 0.385 & 7.611 & 0.634 & 1 & Sm & 13 \\
UGC07151 & 0.747 & 162.726 & 16.273 & 1.090 & 92.254 & 9.225 & 1 & Scd & 11 \\
UGC07232 & 4.692 & 0.391 & 0.130 & 4.677 & 0.321 & 0.107 & 2 & Im & 4 \\
UGC07261 & 0.587 & 29.183 & 4.864 & 0.978 & 17.585 & 2.931 & 2 & Sdm & 7 \\
UGC07323 & 0.565 & 4.591 & 0.510 & 0.866 & 3.180 & 0.353 & 1 & Sdm & 10 \\
UGC07399 & 1.908 & 300.018 & 33.335 & 2.579 & 174.982 & 19.442 & 1 & Sdm & 10 \\
UGC07524 & 0.473 & 52.923 & 1.764 & 0.802 & 10.225 & 0.341 & 1 & Sm & 31 \\
UGC07559 & 0.328 & 2.784 & 0.464 & 0.385 & 1.707 & 0.285 & 2 & Im & 7 \\
UGC07577 & 0.088 & 0.535 & 0.067 & 0.097 & 0.642 & 0.080 & 2 & Im & 9 \\
UGC07603 & 0.903 & 120.260 & 10.933 & 1.218 & 72.781 & 6.616 & 1 & Sd & 12 \\
UGC07608 & 0.719 & 5.802 & 0.829 & 1.010 & 1.756 & 0.251 & 1 & Im & 8 \\
UGC07690 & 0.572 & 48.410 & 8.068 & 0.818 & 38.159 & 6.360 & 2 & Im & 7 \\
UGC07866 & 0.428 & 5.380 & 0.897 & 0.481 & 4.307 & 0.718 & 2 & Im & 7 \\
UGC08286 & 0.786 & 569.823 & 35.614 & 1.250 & 312.461 & 19.529 & 1 & Scd & 17 \\
UGC08490 & 0.683 & 405.417 & 13.980 & 1.106 & 282.014 & 9.725 & 1 & Sm & 30 \\
UGC08550 & 0.485 & 246.394 & 24.639 & 0.754 & 148.222 & 14.822 & 1 & Sd & 11 \\
\end{tabular}
\end{ruledtabular}
\end{table*}

\begin{table*}
\caption{SPARC fits comparison (part 5 of 5)}
\label{SPARC_table_5}
\begin{ruledtabular}
\begin{tabular}{l|ccc|ccc|ccc}
& \multicolumn{3}{c|}{NFW Halo} & \multicolumn{3}{c|}{Perfect Fluid Halo} \\
Galaxy & \begin{tabular}{c}
$\rho_0$ \\
$[\mathrm{GeV/cm^3}]$
\end{tabular} & $\chi^2$ & $\chi^2/\mathrm{dof}$ & \begin{tabular}{c}
$\rho_0$ \\
$[\mathrm{GeV/cm^3}]$
\end{tabular} & $\chi^2$ & $\chi^2/\mathrm{dof}$ & Q & Hubble Type & $N_{data}$\\
\hline
UGC08699 & 2.973 & 522.994 & 13.075 & 3.460 & 135.662 & 3.392 & 2 & Sab & 41 \\
UGC08837 & 0.227 & 4.921 & 0.703 & 0.305 & 10.215 & 1.459 & 2 & Im & 8 \\
UGC09037 & 0.625 & 415.126 & 19.768 & 1.154 & 323.120 & 15.387 & 2 & Scd & 22 \\
UGC09133 & 18.678 & 43435.103 & 648.285 & 7.863 & 3593.666 & 53.637 & 1 & Sab & 68 \\
UGC09992 & 0.165 & 6.673 & 1.668 & 0.241 & 5.310 & 1.328 & 2 & Im & 5 \\
UGC10310 & 0.473 & 24.072 & 4.012 & 0.818 & 9.968 & 1.661 & 1 & Sm & 7 \\
UGC11455 & 3.880 & 1507.863 & 43.082 & 4.564 & 694.647 & 19.847 & 1 & Scd & 36 \\
UGC11557 & 0.119 & 12.776 & 1.161 & 0.241 & 15.733 & 1.430 & 2 & Sdm & 12 \\
UGC11820 & 0.450 & 329.485 & 36.609 & 0.754 & 145.212 & 16.135 & 1 & Sm & 10 \\
UGC11914 & 5.603 & 1313.744 & 20.527 & 6.726 & 1024.021 & 16.000 & 1 & Sab & 65 \\
UGC12506 & 14.314 & 345.037 & 11.501 & 7.751 & 25.611 & 0.854 & 2 & Scd & 31 \\
UGC12632 & 0.406 & 50.828 & 3.631 & 0.722 & 13.522 & 0.966 & 1 & Sm & 15 \\
UGC12732 & 0.666 & 84.249 & 5.617 & 1.026 & 24.092 & 1.606 & 1 & Sm & 16 \\
UGCA281 & 1.860 & 29.389 & 4.898 & 2.003 & 24.802 & 4.134 & 3 & BCD & 7 \\
UGCA442 & 0.292 & 101.638 & 14.520 & 0.561 & 25.083 & 3.583 & 1 & Sm & 8 \\
UGCA444 & 0.595 & 42.508 & 1.215 & 0.657 & 34.084 & 0.974 & 2 & Im & 36 \\
\end{tabular}
\end{ruledtabular}
\end{table*}

\clearpage

\bibliography{referencias}

@article{Nunez:2010ug,
    author = "Nunez, Dario and Gonzalez-Morales, Alma X. and Cervantes-Cota, Jorge L. and Matos, Tonatiuh",
    title = "{Testing DM halos using rotation curves and lensing: A warning on the determination of the halo mass}",
    eprint = "1006.4875",
    archivePrefix = "arXiv",
    primaryClass = "astro-ph.GA",
    doi = "10.1103/PhysRevD.82.024025",
    journal = "Phys. Rev. D",
    volume = "82",
    pages = "024025",
    year = "2010"
}

@ARTICLE{2009ApJ...704.1274W,
       author = {{Walker}, Matthew G. and {Mateo}, Mario and {Olszewski}, Edward W. and {Pe{\~n}arrubia}, Jorge and {Evans}, N. Wyn and {Gilmore}, Gerard},
        title = "{A Universal Mass Profile for Dwarf Spheroidal Galaxies?}",
      journal = {\apj},
         year = 2009,
        month = oct,
       volume = {704},
       number = {2},
        pages = {1274-1287},
          doi = {10.1088/0004-637X/704/2/1274},
archivePrefix = {arXiv},
       eprint = {0906.0341},
 primaryClass = {astro-ph.CO},
       adsurl = {https://ui.adsabs.harvard.edu/abs/2009ApJ...704.1274W}
}

@article{Walker:2007ju,
    author = "Walker, Matthew G. and Mateo, Mario and Olszewski, Edward W. and Gnedin, Oleg Y. and Wang, Xiao and Sen, Bodhisattva and Woodroofe, Michael",
    title = "{Velocity Dispersion Profiles of Seven Dwarf Spheroidal Galaxies}",
    eprint = "0708.0010",
    archivePrefix = "arXiv",
    primaryClass = "astro-ph",
    doi = "10.1086/521998",
    journal = "Astrophys. J. Lett.",
    volume = "667",
    pages = "L53",
    year = "2007"
}

@article{Mateo:2007xh,
    author = "Mateo, Mario and Olszewski, Edward W. and Walker, Matthew G.",
    title = "{The Velocity Dispersion Profile of the Remote Dwarf Spheroidal Galaxy Leo. 1. A Tidal Hit and Run?}",
    eprint = "0708.1327",
    archivePrefix = "arXiv",
    primaryClass = "astro-ph",
    doi = "10.1086/522326",
    journal = "Astrophys. J.",
    volume = "675",
    pages = "201",
    year = "2008"
}

@article{Walker_2009,
doi = {10.1088/0004-6256/137/2/3100},
url = {https://doi.org/10.1088/0004-6256/137/2/3100},
year = {2009},
month = {jan},
publisher = {The American Astronomical Society},
volume = {137},
number = {2},
pages = {3100},
author = {Walker, Matthew G. and Mateo, Mario and Olszewski, Edward W.},
title = {STELLAR VELOCITIES IN THE CARINA, FORNAX, SCULPTOR, AND SEXTANS dSph GALAXIES: DATA FROM THE MAGELLAN/MMFS SURVEY*},
journal = {The Astronomical Journal}
}

@article{Stiele:2010xz,
    author = "Stiele, Rainer and Boeckel, Tillmann and Schaffner-Bielich, Jurgen",
    title = "{Cosmological implications of a Dark Matter self-interaction energy density}",
    eprint = "1003.2304",
    archivePrefix = "arXiv",
    primaryClass = "astro-ph.CO",
    doi = "10.1103/PhysRevD.81.123513",
    journal = "Phys. Rev. D",
    volume = "81",
    pages = "123513",
    year = "2010"
}

@article{Ren:2006tr,
    author = "Ren, Jie and Li, Xue-Qian and Shen, Hong",
    title = "{Equation of state for neutralino star as a form of cold dark matter}",
    eprint = "hep-ph/0604227",
    archivePrefix = "arXiv",
    doi = "10.1088/0253-6102/49/1/43",
    journal = "Commun. Theor. Phys.",
    volume = "49",
    pages = "212--216",
    year = "2008"
}

@article{Shen:1998gq,
    author = "Shen, H. and Toki, H. and Oyamatsu, K. and Sumiyoshi, K.",
    title = "{Relativistic equation of state of nuclear matter for supernova and neutron star}",
    eprint = "nucl-th/9805035",
    archivePrefix = "arXiv",
    doi = "10.1016/S0375-9474(98)00236-X",
    journal = "Nucl. Phys. A",
    volume = "637",
    pages = "435--450",
    year = "1998"
}

@Inbook{Serot1992,
author="Serot, Brian D.
and Walecka, John Dirk",
editor="Ainsworth, T. L.
and Campbell, C. E.
and Clements, B. E.
and Krotscheck, E.",
title="Relativistic Nuclear Many-Body Theory",
bookTitle="Recent Progress in Many-Body Theories: Volume 3",
year="1992",
publisher="Springer US",
address="Boston, MA",
pages="49--92",
isbn="978-1-4615-3466-2",
doi="10.1007/978-1-4615-3466-2_5",
url="https://doi.org/10.1007/978-1-4615-3466-2_5"
}

@article{Jimenez:2002vy,
    author = "Jimenez, Raul and Verde, Licia and Oh, S. Peng",
    title = "{Dark halo properties from rotation curves}",
    eprint = "astro-ph/0201352",
    archivePrefix = "arXiv",
    doi = "10.1046/j.1365-8711.2003.06165.x",
    journal = "Mon. Not. Roy. Astron. Soc.",
    volume = "339",
    pages = "243",
    year = "2003"
}

@book{Carroll:1009754,
      author        = "Carroll, Bradley W and Ostlie, Dale A",
      title         = "{An introduction to modern astrophysics}",
      publisher     = "Addison-Wesley",
      address       = "San Francisco, CA",
      year          = "2007",
      url           = "https://cds.cern.ch/record/1009754",
}

@article{Pawlowski:2018sys,
    author = "Pawlowski, Marcel S.",
    title = "{The Planes of Satellite Galaxies Problem, Suggested Solutions, and Open Questions}",
    eprint = "1802.02579",
    archivePrefix = "arXiv",
    primaryClass = "astro-ph.GA",
    doi = "10.1142/S0217732318300045",
    journal = "Mod. Phys. Lett. A",
    volume = "33",
    number = "06",
    pages = "1830004",
    year = "2018"
}

@article{McGaugh:2012,
    author = "McGaugh, Stacy",
    title = "{The Baryonic Tully-Fisher Relation of Gas Rich Galaxies as a Test of LCDM and MOND}",
    eprint = "1107.2934",
    archivePrefix = "arXiv",
    primaryClass = "astro-ph.CO",
    doi = "10.1088/0004-6256/143/2/40",
    journal = "Astron. J.",
    volume = "143",
    pages = "40",
    year = "2012"
}

@article{Kleyna:2003zt,
    author = "Kleyna, Jan T. and Wilkinson, Mark I. and Gilmore, Gerard and Evans, N. Wyn",
    title = "{A dynamical fossil in the ursa minor dwarf spheroidal galaxy}",
    eprint = "astro-ph/0304093",
    archivePrefix = "arXiv",
    doi = "10.1086/375807",
    journal = "Astrophys. J. Lett.",
    volume = "588",
    pages = "L21--L24",
    year = "2003",
    note = "[Erratum: Astrophys.J.Lett. 589, L59 (2003), Erratum: Astrophys.J. 589, L59 (2003)]"
}

@article{Turner:2022gvw,
    author = "Turner, Michael S.",
    title = "{The Road to Precision Cosmology}",
    eprint = "2201.04741",
    archivePrefix = "arXiv",
    primaryClass = "astro-ph.CO",
    doi = "10.1146/annurev-nucl-111119-041046",
    month = "1",
    year = "2022"
}

@article{Bullock:2017xww,
    author = "Bullock, James S. and Boylan-Kolchin, Michael",
    title = "{Small-Scale Challenges to the $\Lambda$CDM Paradigm}",
    eprint = "1707.04256",
    archivePrefix = "arXiv",
    primaryClass = "astro-ph.CO",
    doi = "10.1146/annurev-astro-091916-055313",
    journal = "Ann. Rev. Astron. Astrophys.",
    volume = "55",
    pages = "343--387",
    year = "2017"
}

@article{Klypin:1999uc,
    author = "Klypin, Anatoly A. and Kravtsov, Andrey V. and Valenzuela, Octavio and Prada, Francisco",
    title = "{Where are the missing Galactic satellites?}",
    eprint = "astro-ph/9901240",
    archivePrefix = "arXiv",
    doi = "10.1086/307643",
    journal = "Astrophys. J.",
    volume = "522",
    pages = "82--92",
    year = "1999"
}

@article{Carlson:1992fn,
    author = "Carlson, Eric D. and Machacek, Marie E. and Hall, Lawrence J.",
    title = "{Self-interacting dark matter}",
    reportNumber = "HUTP-91-A066, LBL-32016, UCB-92-06, NUB-3042-92-TH",
    doi = "10.1086/171833",
    journal = "Astrophys. J.",
    volume = "398",
    pages = "43--52",
    year = "1992"
}

@article{deLaix:1995vi,
    author = "de Laix, Andrew A. and Scherrer, Robert J. and Schaefer, Robert K.",
    title = "{Constraints of selfinteracting dark matter}",
    eprint = "astro-ph/9502087",
    archivePrefix = "arXiv",
    reportNumber = "OSU-TA-9-94, BA-95-08",
    doi = "10.1086/176322",
    journal = "Astrophys. J.",
    volume = "452",
    pages = "495",
    year = "1995"
}

@article{Firmani:2000ce,
    author = "Firmani, C. and D'Onghia, E. and Avila-Reese, V. and Chincarini, G. and Hernandez, X.",
    title = "{Evidence of self-interacting cold dark matter from galactic to galaxy cluster scales}",
    eprint = "astro-ph/0002376",
    archivePrefix = "arXiv",
    doi = "10.1046/j.1365-8711.2000.03555.x",
    journal = "Mon. Not. Roy. Astron. Soc.",
    volume = "315",
    pages = "L29",
    year = "2000"
}

@article{Spergel:1999mh,
    author = "Spergel, David N. and Steinhardt, Paul J.",
    title = "{Observational evidence for selfinteracting cold dark matter}",
    eprint = "astro-ph/9909386",
    archivePrefix = "arXiv",
    doi = "10.1103/PhysRevLett.84.3760",
    journal = "Phys. Rev. Lett.",
    volume = "84",
    pages = "3760--3763",
    year = "2000"
}

@article{Vogelsberger:2012sa,
    author = "Vogelsberger, Mark and Zavala, Jesus",
    title = "{Direct detection of self-interacting dark matter}",
    eprint = "1211.1377",
    archivePrefix = "arXiv",
    primaryClass = "astro-ph.CO",
    doi = "10.1093/mnras/sts712",
    journal = "Mon. Not. Roy. Astron. Soc.",
    volume = "430",
    pages = "1722--1735",
    year = "2013"
}

@ARTICLE{Markevitch2004ApJ,
       author = {{Markevitch}, M. and {Gonzalez}, A.~H. and {Clowe}, D. and {Vikhlinin}, A. and {Forman}, W. and {Jones}, C. and {Murray}, S. and {Tucker}, W.},
        title = "{Direct Constraints on the Dark Matter Self-Interaction Cross Section from the Merging Galaxy Cluster 1E 0657-56}",
      journal = {\apj},
         year = 2004,
        month = may,
       volume = {606},
       number = {2},
        pages = {819-824},
          doi = {10.1086/383178},
archivePrefix = {arXiv},
       eprint = {astro-ph/0309303},
 primaryClass = {astro-ph},
       adsurl = {https://ui.adsabs.harvard.edu/abs/2004ApJ...606..819M}
}

@article{TULIN20181,
title = {Dark matter self-interactions and small scale structure},
journal = {Physics Reports},
volume = {730},
pages = {1-57},
year = {2018},
note = {Dark matter self-interactions and small scale structure},
issn = {0370-1573},
doi = {https://doi.org/10.1016/j.physrep.2017.11.004},
url = {https://www.sciencedirect.com/science/article/pii/S0370157317304039},
author = {Sean Tulin and Hai-Bo Yu},
}

@article{Kamada:2016euw,
    author = "Kamada, Ayuki and Kaplinghat, Manoj and Pace, Andrew B. and Yu, Hai-Bo",
    title = "{How the Self-Interacting Dark Matter Model Explains the Diverse Galactic Rotation Curves}",
    eprint = "1611.02716",
    archivePrefix = "arXiv",
    primaryClass = "astro-ph.GA",
    doi = "10.1103/PhysRevLett.119.111102",
    journal = "Phys. Rev. Lett.",
    volume = "119",
    number = "11",
    pages = "111102",
    year = "2017"
}

@article{Girmohanta2022,
  title = {Cross section calculations in theories of self-interacting dark matter},
  author = {Girmohanta, Sudhakantha and Shrock, Robert},
  journal = {Phys. Rev. D},
  volume = {106},
  issue = {6},
  pages = {063013},
  numpages = {15},
  year = {2022},
  month = {Sep},
  publisher = {American Physical Society},
  doi = {10.1103/PhysRevD.106.063013},
  url = {https://link.aps.org/doi/10.1103/PhysRevD.106.063013}
}

@article{Ghez:2008ms,
    author = "Ghez, A. M. and others",
    title = "{Measuring Distance and Properties of the Milky Way's Central Supermassive Black Hole with Stellar Orbits}",
    eprint = "0808.2870",
    archivePrefix = "arXiv",
    primaryClass = "astro-ph",
    doi = "10.1086/592738",
    journal = "Astrophys. J.",
    volume = "689",
    pages = "1044--1062",
    year = "2008"
}

@article{PhysRevD,
  title = {Compact stars made of fermionic dark matter},
  author = {Narain, Gaurav and Schaffner-Bielich, J\"urgen and Mishustin, Igor N.},
  journal = {Phys. Rev. D},
  volume = {74},
  issue = {6},
  pages = {063003},
  numpages = {14},
  year = {2006},
  month = {Sep},
  publisher = {American Physical Society},
  doi = {10.1103/PhysRevD.74.063003},
  url = {https://link.aps.org/doi/10.1103/PhysRevD.74.063003}
}

@article{Muller:2004yb,
    author = "Muller, Christian M.",
    title = "{Cosmological bounds on the equation of state of dark matter}",
    eprint = "astro-ph/0410621",
    archivePrefix = "arXiv",
    reportNumber = "HD-THEP-04-44",
    doi = "10.1103/PhysRevD.71.047302",
    journal = "Phys. Rev. D",
    volume = "71",
    pages = "047302",
    year = "2005"
}

@article{Calabrese:2009zza,
    author = "Calabrese, Erminia and Migliaccio, Marina and Pagano, Luca and De Troia, Grazia and Melchiorri, Alessandro and Natoli, Paolo",
    title = "{Cosmological constraints on the matter equation of state}",
    doi = "10.1103/PhysRevD.80.063539",
    journal = "Phys. Rev. D",
    volume = "80",
    pages = "063539",
    year = "2009"
}

@article{Kumar:2012gr,
    author = "Kumar, Suresh and Xu, Lixin",
    title = "{Observational constraints on variable equation of state parameters of dark matter and dark energy after Planck}",
    eprint = "1207.5582",
    archivePrefix = "arXiv",
    primaryClass = "gr-qc",
    doi = "10.1016/j.physletb.2014.08.059",
    journal = "Phys. Lett. B",
    volume = "737",
    pages = "244--247",
    year = "2014"
}

@article{Armendariz-Picon:2013jej,
    author = "Armendariz-Picon, Cristian and Neelakanta, Jayanth T.",
    title = "{How Cold is Cold Dark Matter?}",
    eprint = "1309.6971",
    archivePrefix = "arXiv",
    primaryClass = "astro-ph.CO",
    doi = "10.1088/1475-7516/2014/03/049",
    journal = "JCAP",
    volume = "03",
    pages = "049",
    year = "2014"
}

@article{Xu:2013mqe,
    author = "Xu, Lixin and Chang, Yadong",
    title = "{Equation of State of Dark Matter after Planck Data}",
    eprint = "1310.1532",
    archivePrefix = "arXiv",
    primaryClass = "astro-ph.CO",
    doi = "10.1103/PhysRevD.88.127301",
    journal = "Phys. Rev. D",
    volume = "88",
    pages = "127301",
    year = "2013"
}

@article{Serra:2011jh,
    author = "Serra, Ana Laura and Romero, Mariano Javier de Leon Dominguez",
    title = "{Measuring the dark matter equation of state}",
    eprint = "1103.5465",
    archivePrefix = "arXiv",
    primaryClass = "gr-qc",
    doi = "10.1111/j.1745-3933.2011.01082.x",
    journal = "Mon. Not. Roy. Astron. Soc.",
    volume = "415",
    pages = "74",
    year = "2011"
}

@article{Barranco:2013wy,
    author = "Barranco, Juan and Bernal, Argelia and Nunez, Dario",
    title = "{Dark matter equation of state from rotational curves of galaxies}",
    eprint = "1301.6785",
    archivePrefix = "arXiv",
    primaryClass = "astro-ph.CO",
    doi = "10.1093/mnras/stv302",
    journal = "Mon. Not. Roy. Astron. Soc.",
    volume = "449",
    number = "1",
    pages = "403--413",
    year = "2015"
}

@article{Acena:2021wjx,
    author = "Ace{\~n}a, Andr{\'e}s and Barranco, Juan and Bernal, Argelia and L{\'o}pez, Ericson and Llerena, Mario",
    title = "{Preliminary Study of Dark-matter-dominated Systems: Further Analysis for Galactic Dark Matter Halos with Pressure}",
    eprint = "2112.05865",
    archivePrefix = "arXiv",
    primaryClass = "astro-ph.GA",
    doi = "10.3847/1538-4357/ad5439",
    journal = "Astrophys. J.",
    volume = "970",
    number = "2",
    pages = "186",
    year = "2024"
}

@article{Boshkayev:2021wns,
    author = "Boshkayev, Kuantay and Konysbayev, Talgar and Kurmanov, Ergali and Luongo, Orlando and Malafarina, Daniele and Mutalipova, Kalbinur and Zhumakhanova, Gulnur",
    title = "{Effects of non-vanishing dark matter pressure in the Milky Way Galaxy}",
    eprint = "2107.00138",
    archivePrefix = "arXiv",
    primaryClass = "astro-ph.GA",
    doi = "10.1093/mnras/stab2571",
    journal = "Mon. Not. Roy. Astron. Soc.",
    volume = "508",
    number = "1",
    pages = "1543--1554",
    year = "2021"
}

@article{Randall2017,
author = {Randall, Lisa and Scholtz, Jakub and Unwin, James},
doi = {10.1093/mnras/stx161},
journal = {Mon.Not.Roy.Astron.Soc.},
pages = {1515--1525},
title = {{Cores in dwarf galaxies from Fermi repulsion}},
volume = {467},
year = {2017}
}

@article{Destri2013,
archivePrefix = {arXiv},
arxivId = {arXiv:1204.3090v3},
author = {Destri, C and Vega, H J De and Sanchez, N G},
eprint = {arXiv:1204.3090v3},
journal = {New Astronomy},
pages = {39--50},
title = {{Fermionic warm dark matter produces galaxy cores in the observed scales because of quantum mechanics}},
volume = {22},
year = {2013}
}

@article{Destri2013a,
archivePrefix = {arXiv},
arxivId = {arXiv:1301.1864v3},
author = {Destri, C and Vega, H J De and Sanchez, N G},
eprint = {arXiv:1301.1864v3},
journal = {Astroparticle Physics},
pages = {14--22},
title = {{Quantum WDM fermions and gravitation determine the observed galaxy structures}},
volume = {46},
year = {2013}
}

@article{Domcke2015,
author = {Domcke, Valiere and Urbano, Alfredo},
doi = {10.1088/1475-7516/2015/01/002},
journal = {Journal of Cosmology and Astroparticle Physics},
title = {{Dwarf spheroidal galaxies as degenerate gas of free fermions}},
volume = {2015},
year = {2015}
}

@misc{barranco2019constraining,
      title={Constraining ultra light fermionic dark matter with Milky Way observations}, 
      author={J. Barranco and A. Bernal and D. Delepine},
      year={2019},
      eprint={1811.11125},
      archivePrefix={arXiv},
      primaryClass={hep-ph}
}

@inproceedings{deVega:2013ysa,
    author = "de Vega, H. J. and Sanchez, N. G.",
    title = "{Dark matter in galaxies: the dark matter particle mass is about 7 keV}",
    eprint = "1304.0759",
    archivePrefix = "arXiv",
    primaryClass = "astro-ph.CO",
    month = "4",
    year = "2013"
}

@article{Chavanis2022,
  title = {Predictive model of fermionic dark matter halos with a quantum core and an isothermal atmosphere},
  author = {Chavanis, Pierre-Henri},
  journal = {Phys. Rev. D},
  volume = {106},
  issue = {4},
  pages = {043538},
  numpages = {46},
  year = {2022},
  month = {Aug},
  publisher = {American Physical Society},
  doi = {10.1103/PhysRevD.106.043538},
  url = {https://link.aps.org/doi/10.1103/PhysRevD.106.043538}
}

@article{chavanis:hal-01094150,
  TITLE = {{Models of dark matter halos based on statistical mechanics: I. The classical King model}},
  AUTHOR = {Chavanis, Pierre-Henri and Lemou, Mohammed and M{\'e}hats, Florian},
  URL = {https://hal.science/hal-01094150},
  JOURNAL = {{Physical Review D}},
  PUBLISHER = {{American Physical Society}},
  VOLUME = {91},
  NUMBER = {6},
  PAGES = {30},
  YEAR = {2015},
  DOI = {10.1103/PhysRevD.91.063531},
  HAL_ID = {hal-01094150},
  HAL_VERSION = {v1},
}

@article{chavanis:hal-00139138,
  TITLE = {{Phase transitions in self-gravitating systems. Self-gravitating fermions and hard spheres models}},
  AUTHOR = {Chavanis, Pierre-Henri},
  URL = {https://hal.science/hal-00139138},
  NOTE = {New material added},
  JOURNAL = {{Physical Review E : Statistical, Nonlinear, and Soft Matter Physics}},
  PUBLISHER = {{American Physical Society}},
  VOLUME = {65},
  PAGES = {056123},
  YEAR = {2002},
  HAL_ID = {hal-00139138},
  HAL_VERSION = {v1},
}

@article{Arguelles:2023nlh,
    author = {Arg\"uelles, C. R. and Becerra-Vergara, E. A. and Rueda, J. A. and Ruffini, R.},
    title = "{Fermionic Dark Matter: Physics, Astrophysics, and Cosmology}",
    eprint = "2304.06329",
    archivePrefix = "arXiv",
    primaryClass = "astro-ph.GA",
    doi = "10.3390/universe9040197",
    journal = "Universe",
    volume = "9",
    number = "4",
    pages = "197",
    year = "2023"
}

@article{Sanchez2021,
    author = {Sánchez Almeida, Jorge and Trujillo, Ignacio},
    title = "{Numerical simulations of dark matter haloes produce polytropic central cores when reaching thermodynamic equilibrium}",
    journal = {Monthly Notices of the Royal Astronomical Society},
    volume = {504},
    number = {2},
    pages = {2832-2840},
    year = {2021},
    month = {04},
    issn = {0035-8711},
    doi = {10.1093/mnras/stab1103},
    url = {https://doi.org/10.1093/mnras/stab1103},
    eprint = {https://academic.oup.com/mnras/article-pdf/504/2/2832/37789309/stab1103.pdf},
}

@article{Sofue2013,
author = {Sofue, Yoshiaki},
journal = {Publications of the Astronomical Society of Japan},
number = {6},
pages = {118},
title = {{Rotation Curve and Mass Distribution in the Galactic Center — From Black Hole to Entire Galaxy —}},
volume = {65},
year = {2013}
}

@article{deBlok1997,
    author = {de Blok, W. J. G. and McGaugh, S. S.},
    title = "{The dark and visible matter content of low surface brightness disc galaxies}",
    journal = {Monthly Notices of the Royal Astronomical Society},
    volume = {290},
    number = {3},
    pages = {533-552},
    year = {1997},
    month = {09},
    issn = {0035-8711},
    doi = {10.1093/mnras/290.3.533},
    url = {https://doi.org/10.1093/mnras/290.3.533},
    eprint = {https://academic.oup.com/mnras/article-pdf/290/3/533/18540302/290-3-533.pdf},
}

@article{deBlok2001,
    author = "de Blok, W. J. G. and McGaugh, Stacy S. and Rubin, Vera C.",
    title = "{High-Resolution Rotation Curves of Low Surface Brightness Galaxies. II. Mass Models}",
    doi = "10.1086/323450",
    journal = "Astron. J.",
    volume = "122",
    pages = "2396--2427",
    year = "2001"
}

@ARTICLE{McGaugh2001,
       author = {{McGaugh}, Stacy S. and {Rubin}, Vera C. and {de Blok}, W.~J.~G.},
        title = "{High-Resolution Rotation Curves of Low Surface Brightness Galaxies. I. Data}",
      journal = "Astron. J.",
         year = 2001,
        month = nov,
       volume = {122},
       number = {5},
        pages = {2381-2395},
          doi = {10.1086/323448},
archivePrefix = {arXiv},
       eprint = {astro-ph/0107326},
 primaryClass = {astro-ph},
       adsurl = {https://ui.adsabs.harvard.edu/abs/2001AJ....122.2381M}
}

@article{2009Sofue,
    author = "Sofue, Y. and Honma, M. and Omodaka, T.",
    title = "{Unified Rotation Curve of the Galaxy -- Decomposition into de Vaucouleurs Bulge, Disk, Dark Halo, and the 9-kpc Rotation Dip --}",
    eprint = "0811.0859",
    archivePrefix = "arXiv",
    primaryClass = "astro-ph",
    doi = "10.1093/pasj/61.2.227",
    journal = "Publ. Astron. Soc. Jap.",
    volume = "61",
    pages = "227",
    year = "2009"
}

@article{Planck2018,
    author = "Aghanim, N. and others",
    collaboration = "Planck",
    title = "{Planck 2018 results. VI. Cosmological parameters}",
    eprint = "1807.06209",
    archivePrefix = "arXiv",
    primaryClass = "astro-ph.CO",
    doi = "10.1051/0004-6361/201833910",
    journal = "Astron. Astrophys.",
    volume = "641",
    pages = "A6",
    year = "2020",
    note = "[Erratum: Astron.Astrophys. 652, C4 (2021)]"
}

@article{DKoester_1990,
doi = {10.1088/0034-4885/53/7/001},
url = {https://dx.doi.org/10.1088/0034-4885/53/7/001},
year = {1990},
month = {jul},
publisher = {},
volume = {53},
number = {7},
pages = {837},
author = {D Koester and  G Chanmugam},
title = {Physics of white dwarf stars},
journal = {Reports on Progress in Physics},
}

@article{Simon2007,
author = {Simon, Joshua D and Geha, Marla},
journal = {The Astrophysical Journal},
number = {1},
pages = {331--331},
title = {{The Kinematics of the Ultra-faint Milky Way Satellites: Solving the Missing Satellite Problem}},
volume = {670},
year = {2007}
}

@article{Boylan-kolchin2011,
author = {Boylan-Kolchin, Michael and Bullock, James S and Kaplinghat, Manoj},
doi = {10.1111/j.1745-3933.2011.01074.x},
journal = {Mon.Not.Roy.Astron.Soc.},
pages = {40--44},
title = {{Too big to fail ? The puzzling darkness of massive Milky Way subhaloes}},
volume = {44},
year = {2011}
}

@article{El_Zant_2001,
   title={Dark Halos: The Flattening of the Density Cusp by Dynamical Friction},
   volume={560},
   ISSN={1538-4357},
   url={http://dx.doi.org/10.1086/322516},
   DOI={10.1086/322516},
   number={2},
   journal={The Astrophysical Journal},
   publisher={American Astronomical Society},
   author={El‐Zant, Amr and Shlosman, Isaac and Hoffman, Yehuda},
   year={2001},
   month={Oct},
   pages={636–643}
}

@article{Zolotov_2012,
   title={BARYONS MATTER: WHY LUMINOUS SATELLITE GALAXIES HAVE REDUCED CENTRAL MASSES},
   volume={761},
   ISSN={1538-4357},
   url={http://dx.doi.org/10.1088/0004-637X/761/1/71},
   DOI={10.1088/0004-637x/761/1/71},
   number={1},
   journal={The Astrophysical Journal},
   publisher={American Astronomical Society},
   author={Zolotov, Adi and Brooks, Alyson M. and Willman, Beth and Governato, Fabio and Pontzen, Andrew and Christensen, Charlotte and Dekel, Avishai and Quinn, Tom and Shen, Sijing and Wadsley, James},
   year={2012},
   month={Nov},
   pages={71}
}

@article{Brooks_2014,
   title={WHY BARYONS MATTER: THE KINEMATICS OF DWARF SPHEROIDAL SATELLITES},
   volume={786},
   ISSN={1538-4357},
   url={http://dx.doi.org/10.1088/0004-637X/786/2/87},
   DOI={10.1088/0004-637x/786/2/87},
   number={2},
   journal={The Astrophysical Journal},
   publisher={American Astronomical Society},
   author={Brooks, Alyson M. and Zolotov, Adi},
   year={2014},
   month={Apr},
   pages={87}
}

@article{Navarro_1996,
   title={The cores of dwarf galaxy haloes},
   volume={283},
   ISSN={1365-2966},
   url={http://dx.doi.org/10.1093/mnras/283.3.L72},
   DOI={10.1093/mnras/283.3.l72},
   number={3},
   journal={Monthly Notices of the Royal Astronomical Society},
   publisher={Oxford University Press (OUP)},
   author={Navarro, J. F. and Eke, V. R. and Frenk, C. S.},
   year={1996},
   month={Dec},
   pages={L72–L78}
}

@article{Hu2000,
archivePrefix = {arXiv},
arxivId = {arXiv:astro-ph/0003365v2},
author = {Hu, Wayne and Barkana, Rennan and Gruzinov, Andrei},
eprint = {0003365v2},
journal = {Physical Review Letters},
pages = {8--11},
primaryClass = {arXiv:astro-ph},
title = {{Fuzzy Cold Dark Matter: The Wave Properties of Ultralight Particles}},
volume = {85},
year = {2000}
}

@ARTICLE{1997Navarro,
       author = {{Navarro}, Julio F. and {Frenk}, Carlos S. and {White}, Simon D.~M.},
        title = "{A Universal Density Profile from Hierarchical Clustering}",
      journal = {\apj},
         year = 1997,
        month = dec,
       volume = {490},
       number = {2},
        pages = {493-508},
          doi = {10.1086/304888},
archivePrefix = {arXiv},
       eprint = {astro-ph/9611107},
 primaryClass = {astro-ph},
       adsurl = {https://ui.adsabs.harvard.edu/abs/1997ApJ...490..493N}
}

@book{Chandrasekhar1939,
  title     = "An Introduction to the Study of the Stellar Structure",
  author    = "Chandrasekhar, S",
  year      = 1939,
  publisher = "University of Chicago Press",
}

@article{Donato:2009ab,
    author = "Donato, F. and Gentile, G. and Salucci, P. and Martins, C. Frigerio and Wilkinson, M. I. and Gilmore, G. and Grebel, E. K. and Koch, A. and Wyse, R.",
    title = "{A constant dark matter halo surface density in galaxies}",
    eprint = "0904.4054",
    archivePrefix = "arXiv",
    primaryClass = "astro-ph.CO",
    doi = "10.1111/j.1365-2966.2009.15004.x",
    journal = "Mon. Not. Roy. Astron. Soc.",
    volume = "397",
    pages = "1169--1176",
    year = "2009"
}

@article{Gentile:2009bw,
    author = "Gentile, Gianfranco and Famaey, Benoit and Zhao, HongSheng and Salucci, Paolo",
    title = "{Universality of galactic surface densities within one dark halo scale-length}",
    eprint = "0909.5203",
    archivePrefix = "arXiv",
    primaryClass = "astro-ph.CO",
    doi = "10.1038/nature08437",
    journal = "Nature",
    volume = "461",
    pages = "627",
    year = "2009"
}

@article{Eskew_2012,
doi = {10.1088/0004-6256/143/6/139},
url = {https://dx.doi.org/10.1088/0004-6256/143/6/139},
year = {2012},
month = {may},
publisher = {The American Astronomical Society},
volume = {143},
number = {6},
pages = {139},
author = {Eskew, Michael and Zaritsky, Dennis and Meidt, Sharon},
title = {CONVERTING FROM 3.6 AND 4.5μm FLUXES TO STELLAR MASS},
journal = {The Astronomical Journal},
}

@article{Oh_2008,
doi = {10.1088/0004-6256/136/6/2761},
url = {https://dx.doi.org/10.1088/0004-6256/136/6/2761},
year = {2008},
month = {nov},
publisher = {The American Astronomical Society},
volume = {136},
number = {6},
pages = {2761},
author = {Oh, Se-Heon and de Blok, W. J. G. and Walter, Fabian and Brinks, Elias and Kennicutt, Robert C.},
title = {HIGH-RESOLUTION DARK MATTER DENSITY PROFILES OF THINGS DWARF GALAXIES: CORRECTING FOR NONCIRCULAR MOTIONS},
journal = {The Astronomical Journal},
}

@article{Meidt_2014,
doi = {10.1088/0004-637X/788/2/144},
url = {https://dx.doi.org/10.1088/0004-637X/788/2/144},
year = {2014},
month = {jun},
publisher = {The American Astronomical Society},
volume = {788},
number = {2},
pages = {144}
}

@article{Schombert_2018,
    author = {Schombert, James and McGaugh, Stacy and Lelli, Federico},
    title = {The mass-to-light ratios and the star formation histories of disc galaxies},
    journal = {Monthly Notices of the Royal Astronomical Society},
    volume = {483},
    number = {2},
    pages = {1496-1512},
    year = {2018},
    month = {12},
    issn = {0035-8711},
    doi = {10.1093/mnras/sty3223},
    url = {https://doi.org/10.1093/mnras/sty3223},
    eprint = {https://academic.oup.com/mnras/article-pdf/483/2/1496/27089556/sty3223.pdf},
}

@article{Lelli_2016,
doi = {10.3847/0004-6256/152/6/157},
url = {https://dx.doi.org/10.3847/0004-6256/152/6/157},
year = {2016},
month = {nov},
publisher = {The American Astronomical Society},
volume = {152},
number = {6},
pages = {157},
author = {Lelli, Federico and McGaugh, Stacy S. and Schombert, James M.},
title = {SPARC: MASS MODELS FOR 175 DISK GALAXIES WITH SPITZER PHOTOMETRY AND ACCURATE ROTATION CURVES},
journal = {The Astronomical Journal},
}

@article{Schombert_McGaugh_2014, 
title={Stellar Populations and the Star Formation Histories of LSB Galaxies: III. Stellar Population Models}, 
volume={31}, 
DOI={10.1017/pasa.2014.32}, 
journal={Publications of the Astronomical Society of Australia}, 
author={Schombert, James and McGaugh, Stacy}, 
year={2014}, 
pages={e036}}

@article{McGaugh_2016,
  title = {Radial Acceleration Relation in Rotationally Supported Galaxies},
  author = {McGaugh, Stacy S. and Lelli, Federico and Schombert, James M.},
  journal = {Phys. Rev. Lett.},
  volume = {117},
  issue = {20},
  pages = {201101},
  numpages = {6},
  year = {2016},
  month = {Nov},
  publisher = {American Physical Society},
  doi = {10.1103/PhysRevLett.117.201101},
  url = {https://link.aps.org/doi/10.1103/PhysRevLett.117.201101}
}

@ARTICLE{vanAlbada_1985,
       author = {{van Albada}, T.~S. and {Bahcall}, J.~N. and {Begeman}, K. and {Sancisi}, R.},
        title = "{Distribution of dark matter in the spiral galaxy NGC 3198.}",
      journal = {\apj},
         year = 1985,
        month = aug,
       volume = {295},
        pages = {305-313},
          doi = {10.1086/163375},
       adsurl = {https://ui.adsabs.harvard.edu/abs/1985ApJ...295..305V}
}

@BOOK{deVancoleurs1991,
       author = {{de Vaucouleurs}, Gerard and {de Vaucouleurs}, Antoinette and {Corwin}, Jr., Herold G. and {Buta}, Ronald J. and {Paturel}, Georges and {Fouque}, Pascal},
        title = "{Third Reference Catalogue of Bright Galaxies}",
         year = 1991,
       adsurl = {https://ui.adsabs.harvard.edu/abs/1991rc3..book.....D}
}

@ARTICLE{Schombert1992,
       author = {{Schombert}, James M. and {Bothun}, Gregory D. and {Schneider}, Stephen E. and {McGaugh}, Stacy S.},
        title = "{A Catalog of Low Surface Brightness Galaxies. List II}",
      journal = "Astron. J.",
         year = 1992,
        month = apr,
       volume = {103},
        pages = {1107},
          doi = {10.1086/116129},
       adsurl = {https://ui.adsabs.harvard.edu/abs/1992AJ....103.1107S}
}

\end{document}